%% file: main.tex
\RequirePackage[displaymath]{lineno} %
\documentclass[aps,prd,twocolumn,showpacs,amsmath,amssymb]{revtex4-1}
\usepackage{epsfig}
\usepackage{graphicx}%
\usepackage{dcolumn}%
\usepackage{bm}%
\usepackage{ltablex,booktabs}
\usepackage{overpic}
\usepackage{subfigure}
\usepackage{float}
\usepackage{color}
\usepackage{amsmath}
\usepackage{mathcomp}
\usepackage{mathrsfs}
\usepackage{multirow}
\usepackage{rotating}
\usepackage{amssymb}
\usepackage{gensymb}
\usepackage{amsmath}
\usepackage{tabularx}
\usepackage{overpic}
\usepackage{colortbl}
\usepackage{textcomp}
\usepackage{tabu}
\usepackage{xcolor}
\usepackage{makecell}
\usepackage{subfigure}
\usepackage[unicode]{hyperref}
\usepackage{enumitem}%
\input{bes3sym}

\def\pipipi{\pip\pip\pim}

\newcommand{\putat}[3]{\begin{picture}(0,0)(0,0)\put(#1,#2){#3}\end{picture}}
\newcommand{\RNum}[1]{\uppercase\expandafter{\romannumeral #1\relax}}

\begin{document}
\normalsize
\parskip=5pt plus 1pt minus 1pt

\title{ \boldmath Measurement of the absolute branching fraction of the inclusive decay $\Ds\to \pip\pip\pim X$ }
\vspace{-1cm}

\author{\input{authorlist_2022-08-25}}
\begin{abstract}
Using an $\epem$ collision data sample with a total integrated luminosity of $3.19$ fb$^{-1}$
collected with the BESIII detector
at a center-of-mass energy of 4.178~GeV, the branching fraction 
of the inclusive decay of the $\Ds$ meson to final states including at least three charged pions is measured for the first time to be ${\cal B}(\Ds\to\pipipi X) = (32.81 \pm 0.35_{\rm stat} \pm {0.63_{\rm syst}})\%$. In this measurement the charged pions from $\KS$ meson decays are excluded.
The partial branching fractions of $\Ds\to\pipipi X$ are also measured as a function of the $\pip\pip\pim$ invariant mass. 
\end{abstract}
\maketitle

\section{Introduction}
\label{sec:introduction}
Lepton flavor universality (LFU) is a very hot topic in the field of
particle physics. 
Recent reports from several high energy experiments 
studying $B$ meson semileptonic decays of the form $B\to D^{(*)} {\ell}^+ \nu_\ell$ %
 ($\ell = e$, $\mu$, or
$\tau$) suggest tensions with Standard Model based predictions~\cite{hflav}, and have drawn attention from both experimentalists and theorists. 
Among these results,
the LHCb measurement of the ratio of the branching fractions (BFs) of the \Bz meson semileptonic decays, ${\cal R}(D^{*-})\equiv {\cal B}(\Bz\to D^{*-}\tau^+ \nu_\tau)/{\cal B}(\Bz\to D^{*-}\mu^+ \nu_\mu)$, has attracted a lot of attention. The ${\cal R}(D^{*-})$ measurement with the $\tau$
lepton reconstructed from three prongs, {\it i.e.} charged pions ($\tau^+ \to \pipipi X$,
where $X$ stands for all possible particle combinations)~\cite{rdstlhcbprl}, based on a 3~\invfb data sample of $pp$ collisions, 
has one of the smallest statistical uncertainties. However, it is systematically limited and one of the 
limiting systematic uncertainty is associated with the background from $\Bz\to D^{*-} D_s^{+}(X)$, followed by $\Ds\to\pipipi X$ decays. 
In the LHCb measurement, %
the total systematic uncertainty is 9.1\%, out of which 2.5\% is attributed to the ``$\Ds\to 3\pi X$ decay model'' for these background events %
~\cite{rdstlhcbprl}. Here, $3\pi\equiv \pipipi$. 
A better understanding of the $\Ds\to\pipipi X$ decay dynamics and 
of the related $\pipipi$-system kinematics would directly
enhance the sensitivity of the ${\cal R}(D^{*-})$ measurement at LHCb,
 thus becoming more and more important in view of the steady increase of the integrated luminosity collected by the experiment.
Similarly, the measurement of  ${\cal R}(\Lambda_c^+)\equiv {\cal B}(\Lambda_b^0\to \Lambda_c^+ \tau^- \bar{\nu}_\tau)/{\cal B}(\Lambda_b^0\to \Lambda_c^+ \mu^- \bar{\nu}_\mu)$ at LHCb would strongly benefit from the %
knowledge on the $\Ds\to \pipipi X$ decay as input~\cite{LHCb:2022piu}.

\begin{table*}%
    \caption{Major exclusive decay modes that contribute to the inclusive decay of $\Ds\to \pipipi X$, along with the known BFs. The modes with BFs of large uncertainties
    ($\Kp\etapr$, $\pi^{+} \pi^{+} \pi^{+} \pi^{-} \pi^{-} \pi^{0}$ (without contributions from $\etapr/\omega\to\pipi\piz$), %
    etc.) are not shown. None of the charged pions in the $3\pi$ system comes from $\KS$ meson decays. The values of  ${\cal B}({\cal M}\to\pipi X),\,{\cal M}=\eta,\omega,\etapr,\phi$ and ${\cal B}(\tau^+\to\pipipi X)$ are
    estimated by combining all known exclusive decays that contribute~\cite{pdg}. %
    } 
\begin{center}
\begin{tabular}
    {l|c|c|c|c} 
    \hline\hline 
    Decay mode & ${\cal B}$ (\%) & Note & Ref.\\\hline
    $\Ds\to \etapr \rho^{+} $ & $4.6\pm1.2$ & Scaled by ${\cal B}(\etapr\to\pipi X) \sim 80$\% & \cite{pdg,besiii_etaprrhop}\\
    $\Ds\to \pipipi \eta $ & $3.1 \pm 0.2$ & --- & \cite{ds2eta3pi} \\
    $\Ds\to \etapr \pi^{+} $ &  $3.2 \pm 0.2 $  & Scaled by ${\cal B}(\etapr\to\pipi X) \sim 80$\% & \cite{pdg} \\
    $\Ds\to \eta \pip\piz $ & $2.6\pm0.1$ & Scaled by  ${\cal B}(\eta\to\pipi X) \sim 27$\% & \cite{pdg}\\
    $\Ds\to \omega \pi^{+} \pi^{0} $ & $2.5\pm 0.6$ & Scaled by ${\cal B}(\omega\to\pipi X) \sim 91$\%  & \cite{pdg}\\
    $\Ds\to \phi a_1(1260)^+(\to\pipipi)$ & $1.2\pm0.1$ & & \cite{BESIIIkkpipipi} \\
    $\Ds\to K^+K^-\pipipi$ nonresonant & $0.14\pm0.02$ & & \cite{BESIIIkkpipipi} \\
    $\Ds\to \eta\pip $ & $0.46\pm0.03$  & Scaled by  ${\cal B}(\eta\to\pipi X) \sim 27$\%  & \cite{pdg} \\
    $\Ds\to \omega\pip $ & $0.17\pm0.03$  & Scaled by  ${\cal B}(\omega\to\pipi X) \sim 91$\%  & \cite{pdg} \\
    $\Ds\to \omega \pi^{+} \pi^{+} \pi^{-} $ &  $1.6\pm0.5$ & ---& \cite{pdg}\\
    $\Ds\to \pi^{+} \pi^{+} \pi^{-} $ & $1.08 \pm0.04$ & --- & \cite{pdg}\\
    $\Ds\to \phi \pip $  & $0.72\pm0.03$ & Scaled by ${\cal B}(\phi\to\pipi X) \sim 15.6$\% & \cite{BESIIIkkpi}\\
    $\Ds\to \phi \rho^{+} $  & $0.97\pm0.05$ & Scaled by ${\cal B}(\phi\to\pipi X) \sim 15.6$\% & \cite{BESIIIkkpipi0}\\
    $\Ds\to \pi^{+} \pi^{+} \pi^{+} \pi^{-} \pi^{-} $ & $0.79\pm0.08$ & --- & \cite{pdg}\\
    $\Ds\to \tau^{+} \nu_{\tau} $ & $0.72\pm 0.01$ & %
    Scaled by ${\cal B}(\tau^+\to \pipipi X) \sim 13.5$\% & \cite{pdg, BESIIItaunu}\\
    $\Ds\to K^{0} \pi^{+} \pi^{+} \pi^{-} $ & $0.6\pm 0.2$ & $2\times{\cal B}(\Ds\to \KS\pipipi)$ &\cite{pdg} \\
    $\Ds\to \etapr e^{+}\nu_{e}  $ & $0.10\pm 0.01$ & Scaled by ${\cal B}(\etapr\to \pipi\eta(\to\pipi X)) \sim 12$\%&\cite{pdg} \\
$\Ds\to \etapr \mu^{+}\nu_{\mu}  $  & $0.10\pm 0.01$ & Same as above by assuming LFU &\cite{pdg}\\
\hline
    Sum & $24.7 \pm 1.5\phantom{0}$ &  &\\
\hline
\hline
\end{tabular}
\label{tab:exclmodes}
\end{center}
\end{table*}

The inclusive decay $\Ds\to \pipipi X$ has never been studied directly in experiments. 
Table~\ref{tab:exclmodes} lists the BFs of major exclusive $\Ds$ channels with at least three charged pions of $\pipipi$ in the final states 
that have been measured with relatively good precision to date~\cite{pdg,ds2eta3pi,BESIIIkkpipipi,BESIIIkkpi,BESIIIkkpipi0,BESIIItaunu}. 
The decays $\Ds\to\eta^{(\prime)}\pip (X) $ make up about half of the contributions. Measurements on the inclusive decay rate of $\Ds$ into into final states
with at least one combination of $\pipipi$, $\cal B$$(\Ds\to \pipipi X)$, will shed light on yet unobserved decay modes with rich hadronic contents, such as $\Ds\to \etapr \pipipi$. % and $\Ds\to\omega\eta\pip$. %
Charge-conjugate states are implied throughout this paper. 

In this paper, the first measurement of the BF of the inclusive decay of 
$\Ds\to\pipipi X$ based on 3.19~\invfb of $\epem$ collision data collected at a center-of-mass energy ($E_{\rm cm}$) of 4.178~GeV with the BESIII detector in 2016 is presented.
The paper is organized as follows. Section~\ref{sec:dect} introduces the BESIII detector as well as the data and Monte Carlo (MC) simulated samples used in this analysis. Section~\ref{set:meas} gives an overview of the analysis technique. 
Event selection requirements to fully reconstruct $D_s^-$ decays are described in Sec.~\ref{sec:st}, while 
the selection requirements for the 3$\pi$ system at the signal $\Ds$ side and the further analysis based on the selected signal candidates are presented in Sec.~\ref{sec:dt}. The systematic uncertainties on our measurements are evaluated in Sec.~\ref{sec:sys}, and the final results are summarized in Sec.~\ref{sec:CONLUSION}.
\section{Detector and Data Sets}
\label{sec:dect}
The BESIII detector~\cite{ABLIKIM2010345} records the final state particles of symmetric $e^+e^-$
collisions provided by the BEPCII storage ring~\cite{bepcii} in the $E_{\rm cm}$ range from
2.00 to 4.95~GeV. %
BESIII  has  collected  large  data  samples  in  this  energy region~\cite{dataset},
while this analysis uses the entirety of the 3.19~\invfb data sample collected at $E_{\rm cm} = 4.178$~GeV in 2016.
The cylindrical core of the BESIII detector covers 93\% of the full solid angle and consists of a helium-based multilayer drift chamber~(MDC), a plastic scintillator time-of-flight system~(TOF), and a CsI(Tl) electromagnetic calorimeter~(EMC), which are all enclosed in a superconducting solenoidal magnet providing a 1.0~T magnetic field~\cite{besiii_mag}. The solenoid is supported by an octagonal flux-return yoke instrumented with resistive-plate-counter muon identification modules interleaved with steel. %
The MDC reconstructs charged-particle momenta with a resolution of 0.5\% at $1~{\rm GeV}/c$, measuring the characteristic energy loss ($\dedx$) with 6\% precision.
The EMC measures photon energies with a resolution of $2.5\%$ ($5\%$) at $1$~GeV in the barrel (end-cap) region. The time resolution in the TOF barrel region is 68~ps. %
The end-cap TOF system was upgraded in 2015 using multi-gap resistive plate chambers, providing a time resolution of 60~ps~\cite{detector2015}.

Simulated data samples are produced with a {\sc geant4}~\cite{sim} based MC simulation toolkit, 
which includes the geometric description of the BESIII detector and the detector response. %
The simulation also exploits the {\sc kkmc}~\cite{KKMC} generator to take into 
account the beam energy spread and the initial-state radiation in the $\epem$ annihilations. %
The MC simulation samples, referred to as ``inclusive MC simulation'', include the production of $\Ds$ particles via the process
$\epem\to D_s^{(*)\pm} D_s^{\mp}$ (``$D_s$ inclusive MC simulation''), and other open charm processes, as well as 
the production of vector charmonium(-like) states, and the continuum processes incorporated in {\sc kkmc}. 
The known decay modes are modeled with {\sc evtgen}~\cite{EvtGen} using the BFs taken from
the Particle Data Group~\cite{pdg}, and the remaining unknown decays
from the charmonium states %
with {\sc lundcharm}~\cite{LundCharm}. 
Final  state  radiation from  charged  final  state  particles  is
incorporated using {\sc photos}~\cite{FSR}. 
The effective 
luminosities of the inclusive MC simulation samples correspond to 40 times the data luminosity. 
The inclusive MC simulation samples are used to estimate background contributions.
A subset of the $D_s$ inclusive MC simulation sample, with one $\Ds$ meson %
decaying into final state modes including at least three charged pions, is also used as the signal MC simulation sample of $\Ds\to\pipipi X$ 
to determine detection efficiencies.
\section{Analysis Technique}
\label{set:meas}

At $E_{\rm cm}=4.178$~GeV, $D_s^{\pm}$ mesons are produced in pairs predominantly through the process of  $\epem\to D_s^{*\pm}D_s^{\mp}$.
The $D_s^{*\pm}$ mesons then primarily decay into $D_s^{\pm}\gamma$ (93.5\%)
and $D_s^{\pm}\piz$ (5.8\%)~\cite{pdg}.  
The double-tag (DT) technique, first employed by
the Mark III collaboration~\cite{dtref}, is used to determine
the BF for the inclusive decay mode of the $\Ds$ meson. 
We first fully reconstruct a $\Dsm$ meson, named ``single tag (ST)'', 
in one of the hadron tag modes listed in Sec.~\ref{sec:st}~\cite{ds2pn}.
Then in the ST $\Dsm$ sample, we reconstruct the $\Ds\to \pipipi X$
signal in the side recoiling against the $\Dsm$ candidate,
referred to as the DT. 
%The $\Dsm$ meson decays into one of the hadron tag modes listed in Sec.~\ref{sec:st},
%which is later referred to as single tag (ST).
%
The number of ST candidates for a specific 
tag mode $\alpha$ ($N^{\alpha}_{\rm ST}$) 
determined from the fit shown in Fig.~\ref{fig:mDsfit} 
is used to 
obtain the total number of $\epem \to D_s^{(*)\pm}D_s^{\mp}$ events in the data ($N_{\rm tot}$) 
by

\begin{equation}
    N^{\alpha}_{\rm ST} = 2N_{\rm tot}\cdot {\cal B}_{\alpha}\cdot \epsilon^{\alpha}_{\rm ST}\,,\label{eq:ST}
\end{equation}
where ${\cal B}_{\alpha}$ is the BF for the $\Dsm$ tag mode $\alpha$, and $\epsilon^{\alpha}_{\rm ST}$ is the ST detection efficiency.

Then
the signal decay which contains at least three charged pions 
is selected at the recoil (signal) side. The tag side and signal side selection 
criteria are described in detail in Sec.~\ref{sec:st} and Sec.~\ref{sec:dt}, respectively.
To ensure one 3$\pi$ system per signal decay, in the case of more than one $\pi^-$ (two $\pi^+$s) at the signal side, 
we select the $\pi^-$ (two $\pi^+$s) with the largest momentum (largest and second largest momenta), in the laboratory frame. 
The true (reconstructed) $3\pi$ invariant mass $M(\pipipi)$ is calculated based on true (reconstructed)
pion momenta.

Since the reconstruction efficiency is dependent on the kinematics of the $3\pi$ system at 
the signal side, which is not well known, the partial BFs of the inclusive $\Ds\to\pipipi X$ decay are measured %
in intervals of $M(\pipipi)$. 
The efficiency is parametrized as a function of $M(\pipipi)$, and its
dependence on other kinematic variables is later examined in Sec.~\ref{sec:Ppi}.
For tag mode $\alpha$, 
the number of produced DT events and the numbers of observed DT events
are related
in $M(\pipipi)$ intervals
through a detector response matrix that accounts for detector efficiency and detector resolution for $M(\pipipi)$, 

\begin{equation}
    N^{\alpha}_{{\rm obs},i} = \sum_{j} \epsilon^{\alpha}_{ij} N^{\alpha}_{{\rm prod},j}\,,\label{eq:prodobs}
\end{equation}
where $N^{\alpha}_{{\rm prod},j}$ ($N^{\alpha}_{{\rm obs},i}$) is the number of signal events produced (observed) in the $j^{{th}}$ ($i^{th}$) $M(\pipipi)$ interval. The $ij^{th}$ element of the 
efficiency matrix, $\epsilon^{\alpha}_{ij}$, describes the efficiency and migration across the $M(\pipipi)$ intervals. It is calculated as $\epsilon^{\alpha}_{ij}\equiv N^{\alpha}_{{\rm obs},ij}/N^{\alpha}_{{\rm prod},j}$ using simulated signal $\Ds\to\pipipi X$ decays with
$D_s^-$ decaying into the tag mode $\alpha$. 
Here, $N^{\alpha}_{{\rm obs},ij}$ is the number of signal MC simulated events produced in the $j^{th}$ 
interval and reconstructed in the $i^{th}$ interval of the $M(\pipipi)$ distribution. 

The number of observed $\Ds\to\pipipi X$ signal candidates in each $M(\pipipi)$ interval is determined by fitting the invariant mass distribution of 
ST $D_s^-$ candidates to extract the raw signal yield $N^{\alpha}_{{\rm raw},i}$. %
The yields of expected $\KS$ background contributions, $N^{\alpha}_{{\KS},i}$,
and contributions from particle misidentification (misID), $N^{\alpha}_{{\rm misID},i}$, estimated from MC simulation, are subtracted from the raw yield $N^{\alpha}_{{\rm raw},i}$

\begin{equation}
 N^{\alpha}_{{\rm obs},i} =  N^{\alpha}_{{\rm raw},i} - N^{\alpha}_{{\KS},i} - N^{\alpha}_{{\rm misID},i}.\label{eq:raw2obs}
\end{equation}
Here, $N^{\alpha}_{{\rm misID},i} = N^{\alpha}_{e\to\pi,i} + N^{\alpha}_{\mu\to\pi,i} + N^{\alpha}_{K\to\pi,i}$ 
includes contributions from the three misID scenarios described in Sec.~\ref{sec:pidbkg}.

For tag mode $\alpha$, the number of $\Ds\to\pipipi X$ signal candidates produced in the $i^{th}$ interval of the $M(\pipipi)$ distribution
is then obtained by solving Eq.~(\ref{eq:prodobs}), which gives

\begin{equation}
    N^{\alpha}_{{\rm prod},i} = \sum_{j} (\epsilon^{-1})^{\alpha}_{ij} N^{\alpha}_{{\rm obs},j}\,,\label{eq:solvmat}
\end{equation}
and its statistical uncertainty is given by

\begin{equation}
    \left(\sigma_{\rm stat}\left(N^{\alpha}_{{\rm prod},i}\right)\right)^2 = \sum_{j} \left(\left(\epsilon^{-1}\right)^{\alpha}_{ij}\right)^2 \left(\sigma_{\rm stat}\left(N^{\alpha}_{{\rm obs},j}\right)\right)^2\,,\label{eq:solvmaterr}
\end{equation}
where $\sigma_{\rm stat}\left(N^{\alpha}_{{\rm obs},j}\right)$ is the statistical uncertainty of $N^{\alpha}_{{\rm obs},j}$. The statistical uncertainties of $\epsilon^{\alpha}_{ij}$ due to the limited size of the signal MC simulation sample are later considered as a source of systematic uncertainties, as discussed in Sec.~\ref{sec:mcstat}.

Taking $N_{\rm tot}$ from Eq.~(\ref{eq:ST}), the corresponding partial BF of $\Ds\to\pipipi X$ is given by

\begin{align}
    {\Delta {\cal B}}^{\alpha}_{3\pi X,i}  = \frac{N^{\alpha}_{{\rm prod},i}}{2N_{\rm tot}\cdot {\cal B}_{\alpha}}
    = \frac{N^{\alpha}_{{\rm prod},i}}{  N^{\alpha}_{\rm ST}}\cdot\epsilon^{\alpha}_{\rm ST}\,.\label{eq:pbf}
\end{align}

In each tag mode $\alpha$, the partial BFs of  the $\Ds\to\pipipi X$ decays are summed to obtain the total BF ${\cal B}^\alpha(\Ds\to\pipipi X)  = \sum_i \Delta {\cal B}^\alpha_{3\pi X,i}$. The total BF and its associated statistical uncertainty are 

\begin{align}
     {\cal B}\left(\Ds\to\pipipi X\right) &= \sum_\alpha \frac{{\cal B}^\alpha}{\sigma^2_{{\cal B}^\alpha}} / \sum_\alpha \frac{1}{\sigma^2_{{\cal B}^\alpha}}, 
\end{align}
and
\begin{align}
     \sigma^2_{\rm stat}\left({\cal B}\left(\Ds\to\pipipi X\right)\right) & = 1/\sum_\alpha \frac{1}{\sigma^2_{{\cal B}^\alpha}},
    \label{eq:tbf}
\end{align}
respectively, where $\sigma_{{\cal B}^\alpha}$ is the statistical uncertainty of ${\cal B}^\alpha(\Ds\to\pipipi X)$.

\section{Single Tag Analysis}
\label{sec:st}

ST $\Dsm$ candidates are selected in two hadronic tag modes, $\Dsm\to \KS\Km$ and $\Dsm\to \Km\Kp\pim$. Other tag modes are not considered due to their relatively high background level. %
All charged track candidates %
detected in the MDC 
must be within a polar angle ($\theta$) range of  $\vert\!\cos\theta\vert<0.93$, where $\theta$ 
is defined with respect to the $z$ axis, which is the symmetry axis of the MDC. For charged tracks not originating from
$\KS$ decays, the distance of closest 
approach to the interaction point (IP) is required to be less than 10 cm along the $z$ axis, and less than 1 cm in the 
transverse plane. %

Charged tracks are identified as pions or kaons with the particle identification algorithm (PID),
which combines measurements of \dedx in the MDC
and the time of flight in the TOF to form likelihoods ${\cal L}(h) (h=K,\pi)$
for different hadron $h$ hypotheses. Charged kaons and pions are identified by comparing the likelihoods for the kaon and pion hypotheses,
${\cal L}(K) > {\cal L}(\pi)$ and ${\cal L}(\pi) > {\cal L}(K)$, respectively. 
Furthermore, charged pion candidates with momenta below 100~\mevc\ are rejected to suppress
soft pions from $\Dstarp\to\Dz\pip$. 

The $\KS$ candidates are reconstructed from two oppositely charged tracks %
each with the distance of closest 
approach to the IP less than 20 cm along the $z$ axis.
The two charged tracks are assigned as
$\pipi$ 
without imposing further PID criteria. 
They are constrained to originate from a common vertex and are required to
have an invariant mass satisfying $\vert M(\pipi) - m_{\KS}\vert < 12$~\mevcc, 
where $M(\pipi)$ and $m_{\KS}$ are the invariant mass of the pion pair %
and the known $\KS$ mass~\cite{pdg}, respectively. 

For each $D_s^-$ candidate, the reconstructed invariant
mass $M(D_s^-)$ is required to be in the range 1.90 and 2.05~$\gevcc$. The recoil mass of the $\Dsm$ candidate is defined as

\begin{align}
M_{\rm rec} c^2 \equiv \sqrt{\left(E_{\rm cm} - \sqrt{\left|\vec{p}_{D_s}c\right|^2+\left(m_{D_s}c^2\right)^2}\right)^2-\left|\vec{p}_{D_s}c\right|^2}\,,
\end{align}
where %
$\vec{p}_{D_s}$ is the reconstructed 
momentum of the $\Dsm$ candidate in the $\epem$
center-of-mass frame, and $m_{D_s}$ is the known \Dsm\ mass~\cite{pdg}. 

Furthermore, for the tag mode $\Dsm\to \KS\Km$, background contributions from $\Dz\to\Km\pip$ decays are suppressed by rejecting $\Dsm\to\KS\Km$ candidates with $|M(\Km\pi^+_{K_S})-m_{\Dm}|<10 $~\mevcc, 
where $\pi^+_{K_S}$ is from $\KS\to\pipi$, and $m_{\Dm}$ is the known $\Dm$ mass~\cite{pdg}. %

Finally, for either tag mode, in about 6\% of the cases, more than one $D_s^-$ candidate per event passes the tag side event selection requirements.
In these cases, only the $D_s^-$ candidate with $M_{\rm rec}$ closest to the known $\Dss$ mass per event per tag mode is retained for further study. The systematic uncertainties related to this single-candidate requirement are discussed in Sec.~\ref{sec:SCS}.

\begin{figure}[!htb]
\centering
\subfigure{
\includegraphics[trim=0 77 0 0,clip,width=0.8\linewidth]{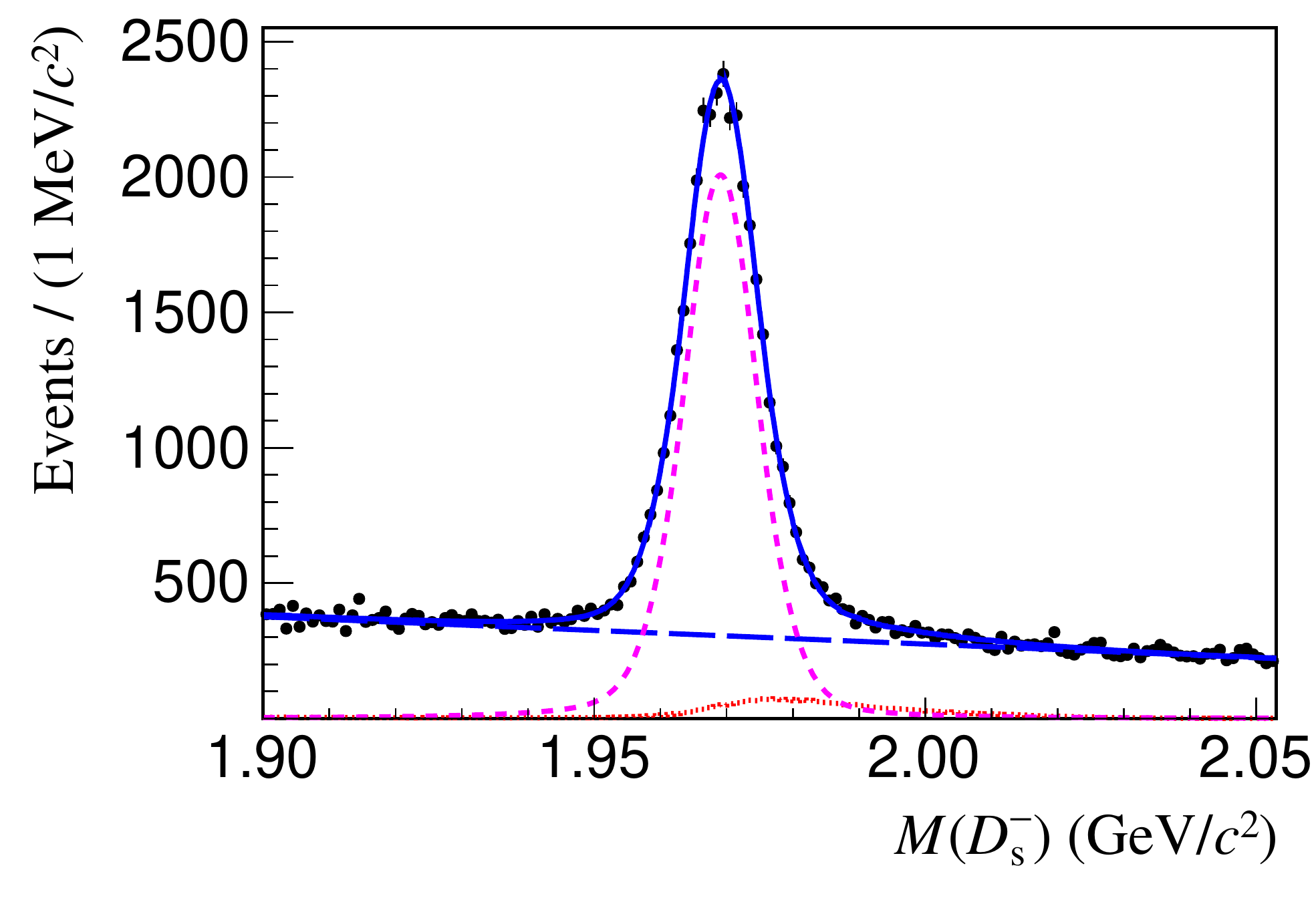}
    \putat{-80}{+74}{$\Dsm\to\KS\Km$}}\vspace{-0.38cm}
    \subfigure{
\includegraphics[trim=0 0 0 12,clip,width=0.8\linewidth]{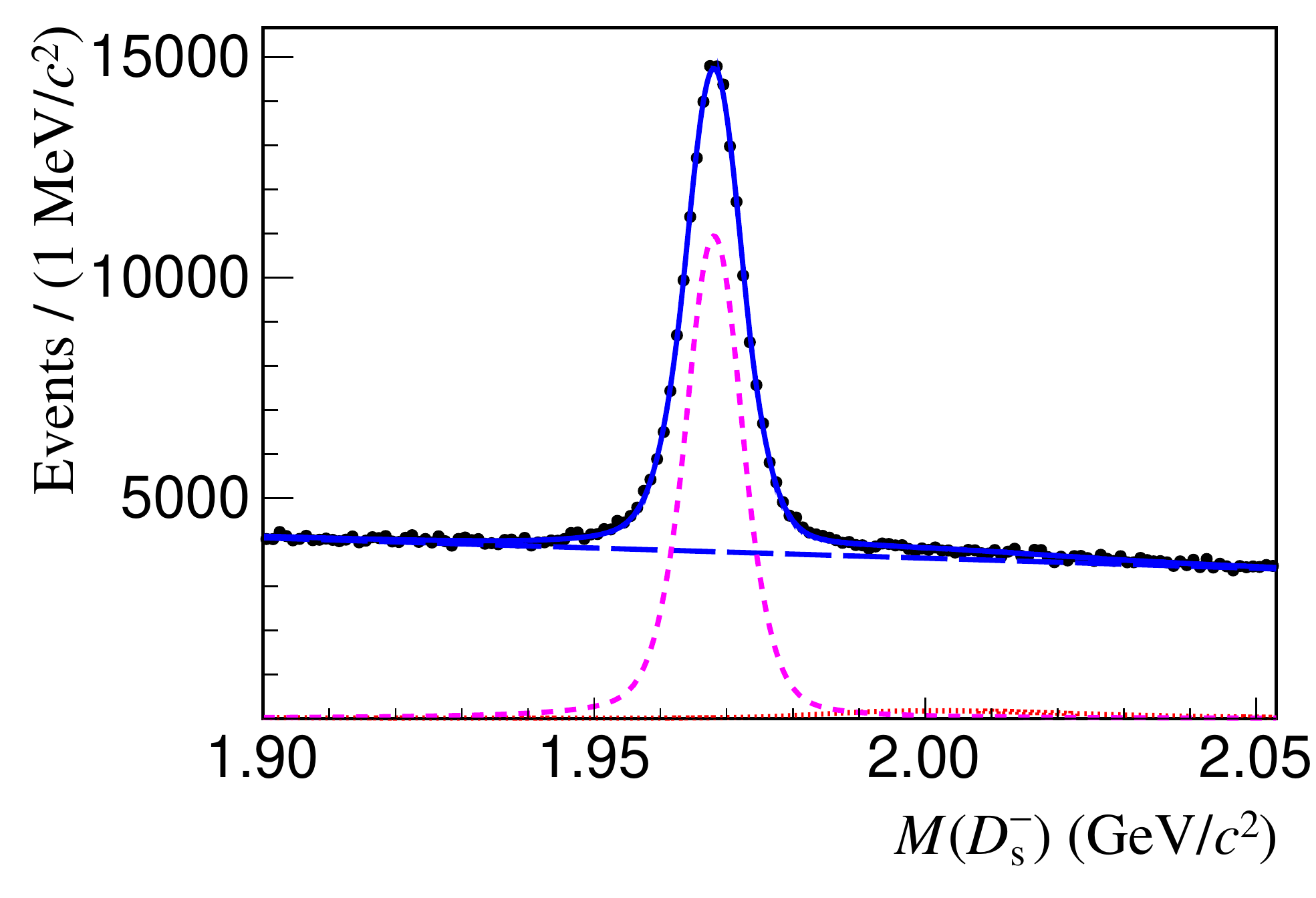}
    \putat{-80}{+102}{$\Dsm\to\Km\Kp\pim$}}

    \caption{The invariant mass $M(\Dsm)$ distributions %
    for the ST modes $\Dsm\to\KS\Km$ (top) and $\Dsm\to\Km\Kp\pim$ (bottom). Data (black points) are shown
    overlaid with the fit results including the total (solid blue),
signal PDF (dashed magenta), $\Dm$ background PDF (dotted red), and combinatorial background PDF (long-dashed blue) components. %
    }
\label{fig:mDsfit}
\end{figure}

For each tag mode, the number of correctly reconstructed ST $\Dsm$ candidates 
is determined
by performing an unbinned maximum likelihood fit to the distribution of $M(\Dsm)$, as shown in Fig.~\ref{fig:mDsfit}. 
The combinatorial background is modeled 
using a linear probability density function (PDF). 
The shapes of the $\Dm$ background contributions from $\Dm\to\KS\pim$ ($\Dm\to\Kp\pim\pim$) decays in the
tag mode $\Dsm\to\KS\Km$ ($\Dsm\to \Km \Kp\pim$) when $\Km$ is misidentified as $\pim$ are modeled with MC simulation. 
The yields of the $\Dm$ meson background contributions are estimated with the simulation and fixed in the fits.
The signal lineshapes are modeled by the sum of a Gaussian PDF and a double-sided Crystal-ball (DSCB) PDF~\cite{dscb} with a common
mean value $m_0$. The Gaussian PDF has a width of $\sigma_0$ that is allowed to float along with $m_0$. 
The DSCB PDF tail parameters and the width ratio between the DSCB and Gaussian PDFs are determined from signal MC simulated events 
and fixed in the fits to data, as in Ref.~\cite{ds23pi}. %
The fitted yields are summarized in Table~\ref{tab:tagyields}, along with the ST efficiencies.

\begin{table}[htbp]
\caption{The ST yields and efficiencies for the two tag modes.}
\begin{center}
\begin{tabular}
    {l|cc} \hline \hline
    Tag mode  & $N_{\rm ST}^\alpha$ & $\epsilon_{\rm ST}^\alpha$%
    \\\hline
    $\Dsm\to\KS\Km$ & $\phantom{0}30673\pm 230$ &  46.3\% \\
    $\Dsm\to\Km\Kp\pim$ &  $134370\pm 565$ & 38.8\% \\\hline \hline
\end{tabular}
\label{tab:tagyields}
\end{center}
\end{table}

\section{Double Tag Analysis}
\label{sec:dt}
In each event, once an ST $\Dsm$ candidate is found, 
three charged pions with the sum of charges opposite to the charge of the $\Dsm$ tag in the rest of the 
event (recoil side) are searched for. The tracking and PID requirements of charged pions are the same as those in the tag side as described in Sec.~\ref{sec:st}.%

\begin{figure*}[!htb]
\centering
\includegraphics[width=0.32\linewidth]{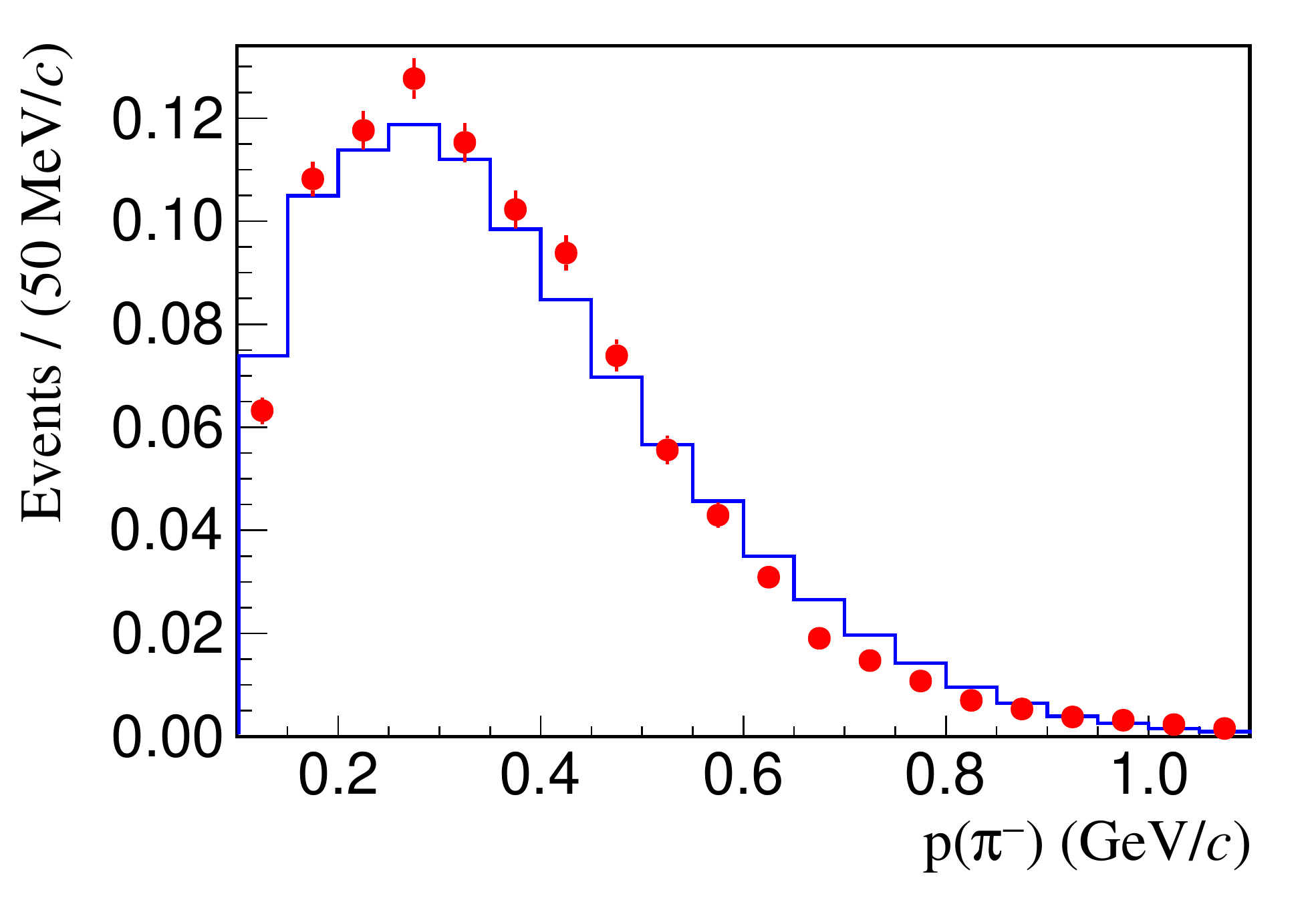}
\includegraphics[width=0.32\linewidth]{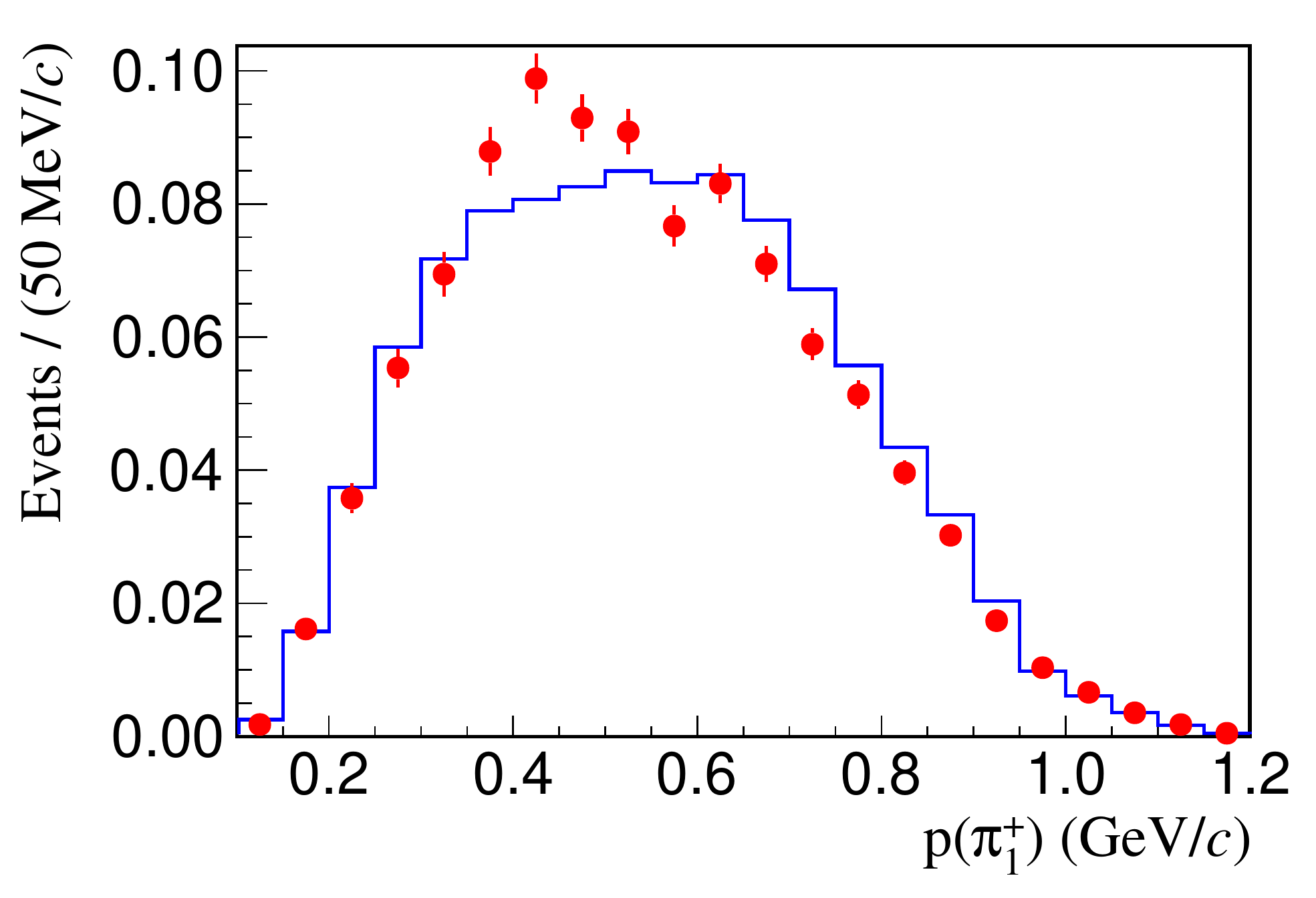}
\includegraphics[width=0.32\linewidth]{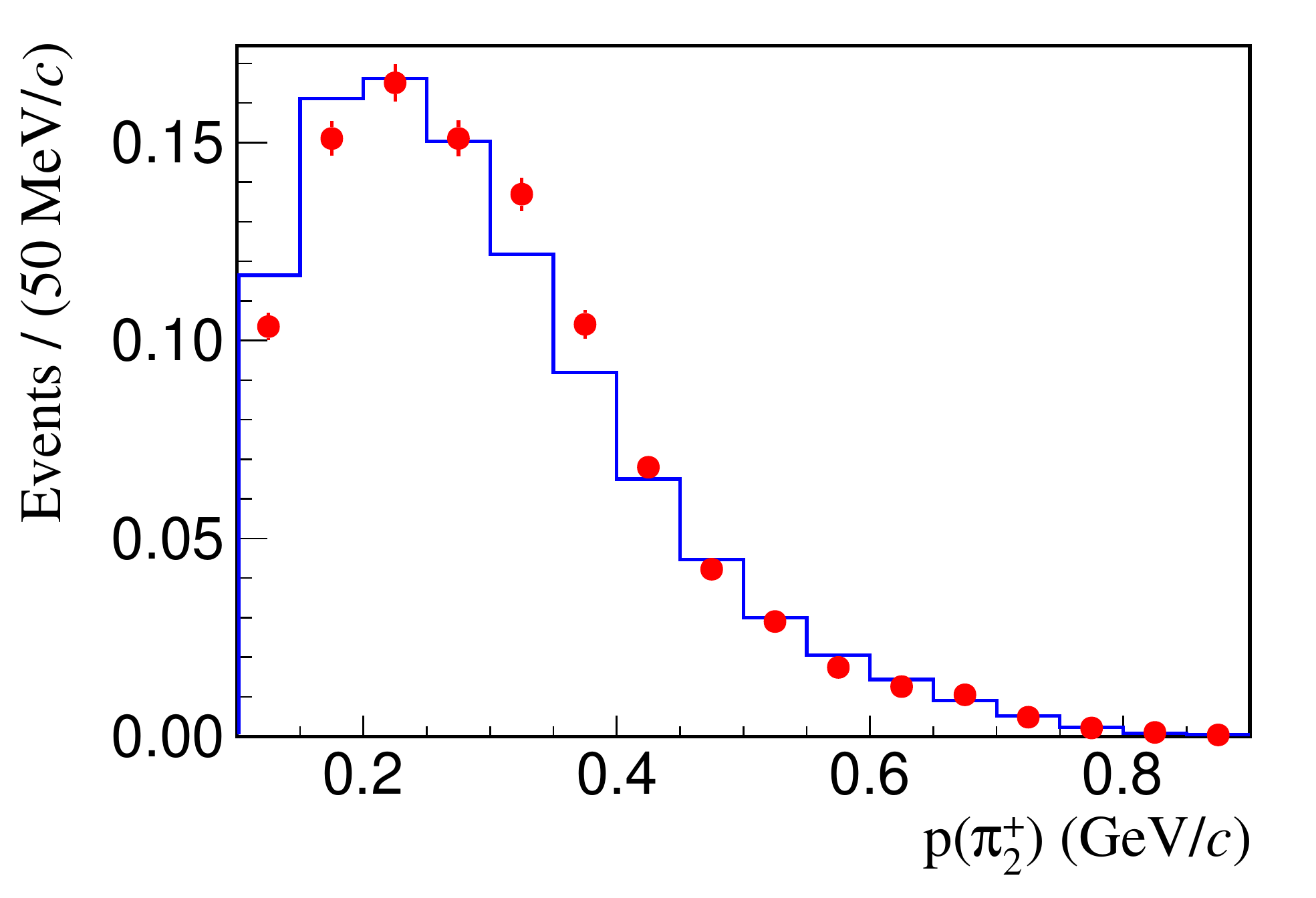}
    \caption{Normalized momentum spectra after background subtraction from the $3\pi$ system of the inclusive decay $\Ds\to\pipipi X$ for the tag mode $\Dsm\to \Km\Kp\pim$. The points are obtained from data and the solid line histograms are from inclusive MC simulation samples.  %
    The subscripts for $\pip$s indicate the order $p(\pi_1^+)>p(\pi_2^+)$. }
\label{fig:Ppi3}
\end{figure*}

\begin{figure*}[!htb]
\centering
\includegraphics[width=0.32\linewidth]{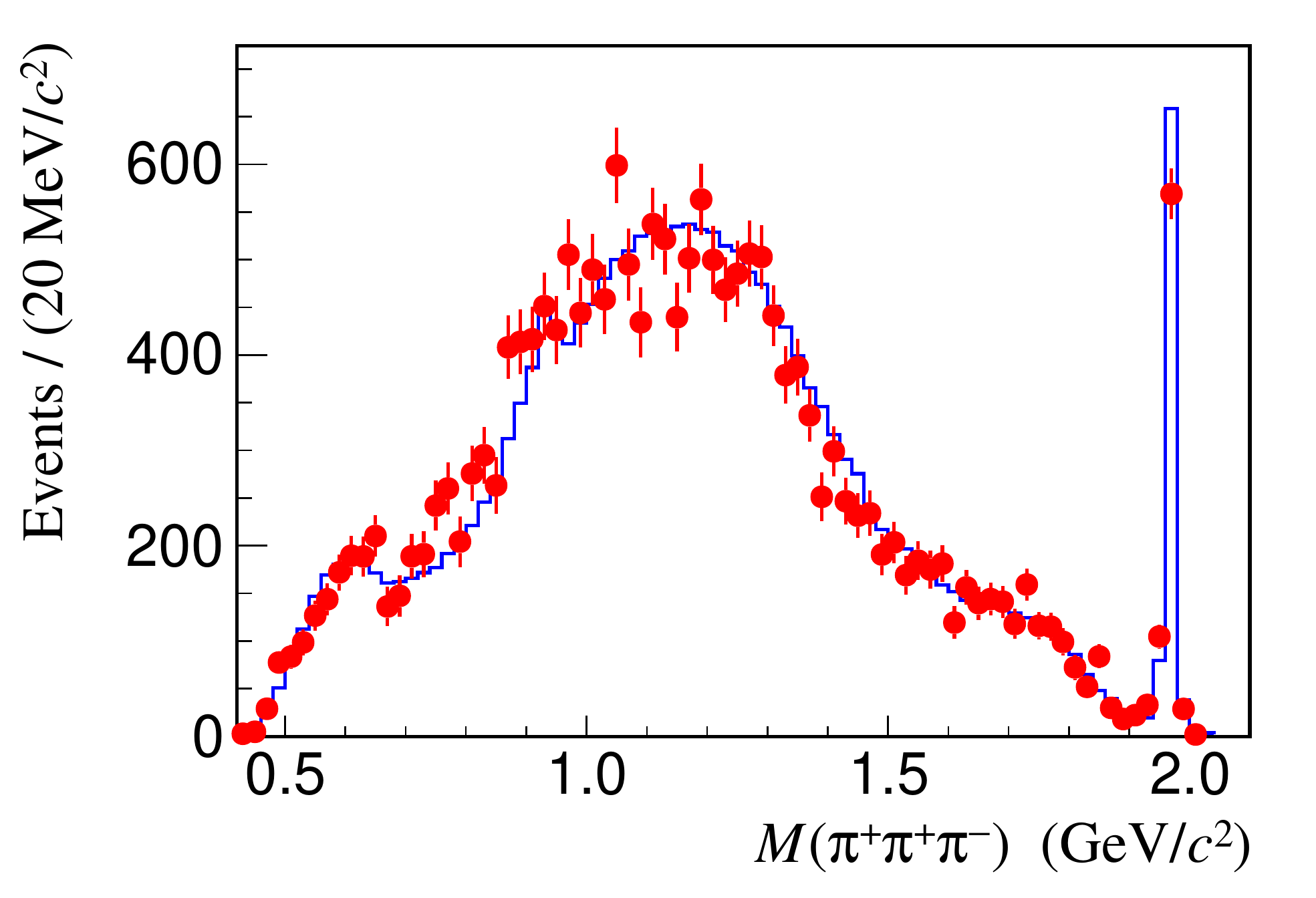}
\includegraphics[width=0.32\linewidth]{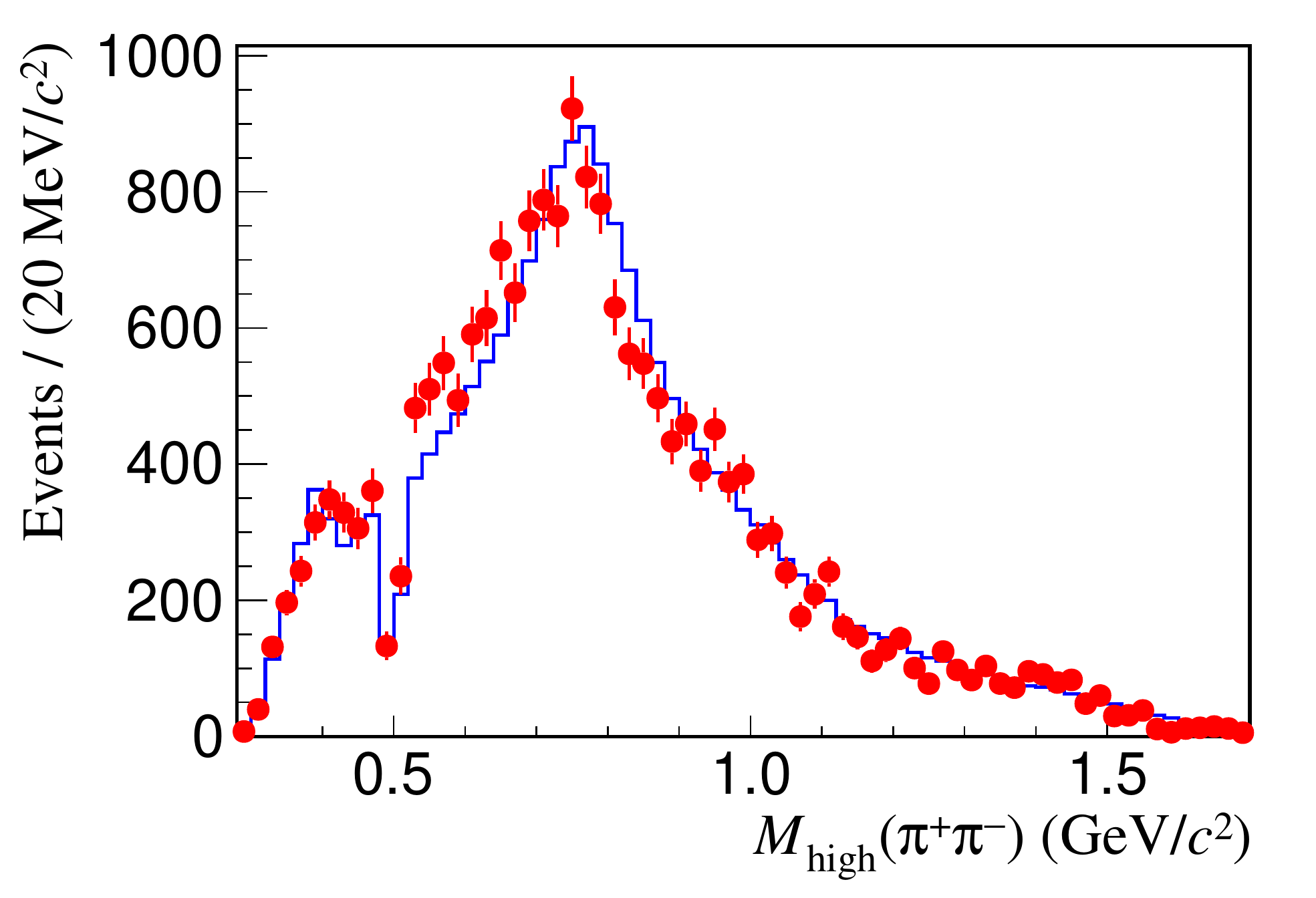}
\includegraphics[width=0.32\linewidth]{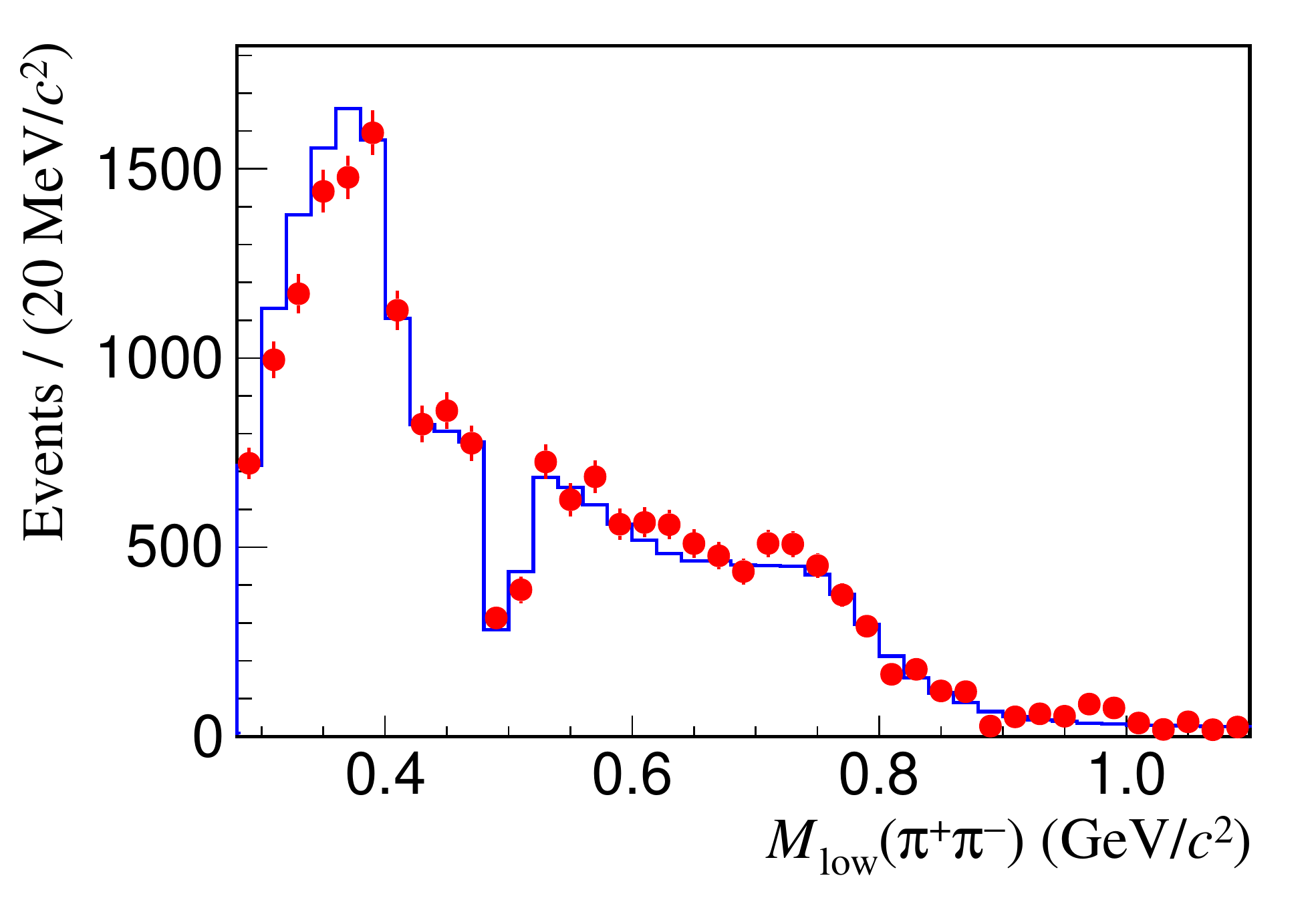}
    \caption{Normalized invariant mass spectra after background subtraction from the $3\pi$ system of the inclusive decay $\Ds\to\pipipi X$ for the tag mode $\Dsm\to \Km\Kp\pim$. The points are obtained from data and the solid line histograms are from inclusive MC simulation samples.  %
    The subscripts for the two $\pipi$ combinations indicate the order $M_{\rm high}(\pipi)>M_{\rm low}(\pipi)$. 
    The dips in the $M(\pipi)$ spectra are due to the 10~\mevcc $\KS$ mass veto defined in Sec.~\ref{sec:ksbkg}.}
\label{fig:M3pi}
\end{figure*}

There are two major background sources that contribute to the $3\pi$ inclusive final states. The first source of background contribution is due to events with three correctly identified charged pions where at least one of them comes from $\KS\to\pipi$ decays. The other type of background comes from events for which at least one charged pion is misidentified.  These two types of background sources are studied separately using dedicated MC simulations of inclusive $\Ds$ decays to determine additional selection criteria, 
and to estimate the remaining contribution after imposing all the requirements. The background subtracted pion momentum spectra and mass spectra from the $3\pi$ system for the tag mode $\Dsm\to \Km\Kp\pim$ are shown in Figs.~\ref{fig:Ppi3} and~\ref{fig:M3pi}, while the spectra for the tag mode $\Dsm\to\KS\Km$ follow very similar patterns with lower yields. 
There is reasonable agreement between data and the simulated sample, and any
differences in the pion momentum spectra are later considered as a source of systematic uncertainties. 

\subsection{\boldmath Background from $\KS$ decays}
\label{sec:ksbkg}

To suppress the background contribution due to $\KS$ decays, each of the three charged pions of the $3\pi$ system is checked against a list of $\KS$ candidates in an event. Each of the $\KS$ candidates is reconstructed 
with the criteria defined in Sec.~\ref{sec:st}
. If at least one pion is part of one reconstructed $\KS$ candidate that has $L/\sigma_L > 2$ ($L$ is defined as the distance between the $\KS$ vertex and the IP, and $\sigma_L$ is the uncertainty of $L$), 
or 
within the mass window $|M(\pipi)-m_{\KS}|<10 $\mevcc, this signal candidate is rejected. 
With this requirement, about 88\% of the $\KS$ meson related background contributions are rejected, while losing about 11\% of the signal events. 

\subsection{Particle misidentification background}
\label{sec:pidbkg}
For one charged pion, there are three possible misID scenarios:
\begin{itemize}
    \item $\mu^{\pm}\to\pi^{\pm}$,
    \item $e^{\pm}\to\pi^{\pm}$,
    \item $K^{\pm}\to\pi^{\pm}$.
\end{itemize}
As the charged pions are identified against kaons, the main misID background contributions are from the semileptonic $\Ds$ decays, %
and the misidentified pion in the $3\pi$ system has the same charge as the signal $\Ds$ side. Electron background can be largely removed by rejecting signal candidates for which at least one $e^+$ has been identified at the signal $\Ds$ side. 
The electron PID is based on 
the likelihoods for the electron, pion, and kaon hypotheses ${\cal L}(h) (h=e,\pi,K)$
using information from the MDC, TOF, and EMC, as in Ref.~\cite{BESIIIDs2eX}.
With this, about 90\% of the $\ep\to\pip$ misID background contributions are removed at the cost of mere $\sim$5\% of the signal events. 
The residual misID backgrounds are then dominated by
the semimuonic decays $\Ds\to X\mup \nu_\mu$, and are estimated from the $D_s$ inclusive MC simulation sample ($\sim$70\%), 
, and the related systematic uncertainties are discussed in Sec.~\ref{sec:pidbkg}.

\subsection{Partial branching fractions}
\label{sec:pBF}
The $\Ds\to\pipipi X$ partial BFs are measured in eleven $M(\pipipi)$ intervals. The intervals are populated with approximately equal numbers of DT candidates, except for the last interval,
where the signal candidates are predominantly from the exclusive decay of $\Ds\to\pipipi$. 
The interval boundaries in the $M(\pipipi)$ distribution are chosen as [0.40, 0.74, 0.90, 0.99, 1.08, 1.15, 1.23, 1.30, 1.41, 1.57, 1.90, 2.05]~\gevcc.

\begin{figure}[!htb]
\centering
\subfigure{
\includegraphics[trim=0 77 0 0,clip,width=0.8\linewidth]{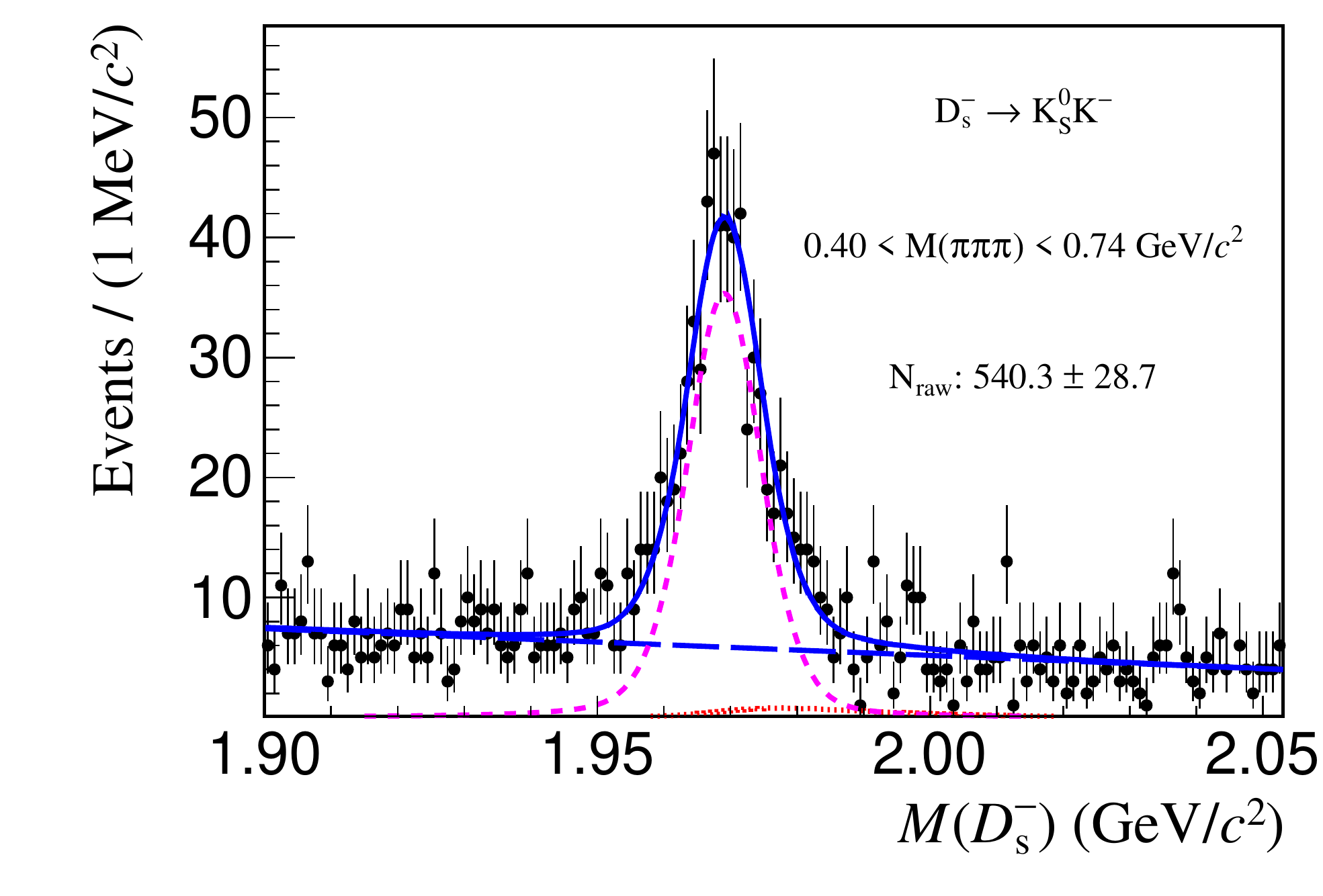}
    \putat{-85}{+89}{\colorbox{white}{$\Dsm\to\KS\Km$}}}\vspace{-0.37cm}
    \subfigure{
\includegraphics[trim=0 0 0 12,clip,width=0.8\linewidth]{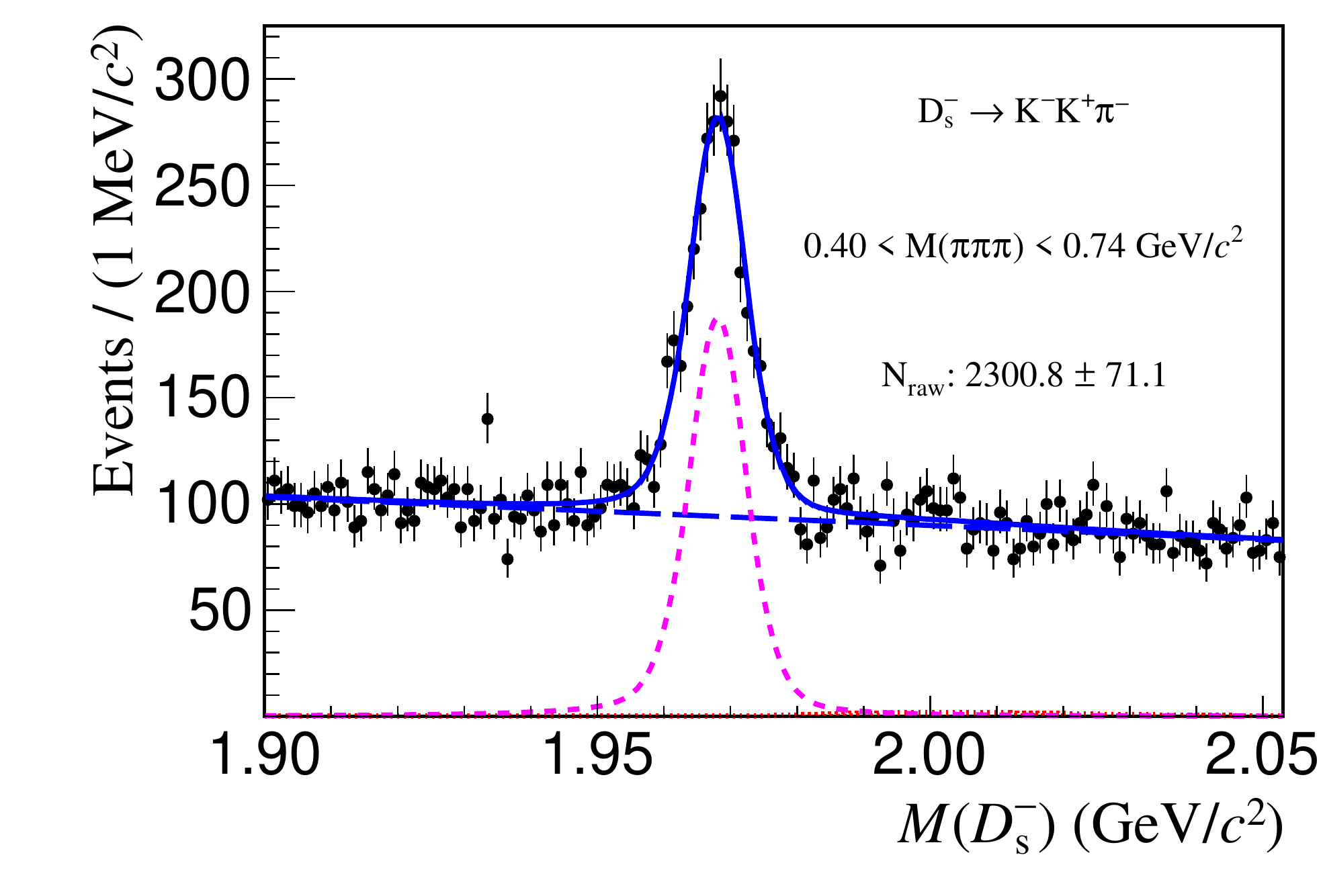}
    \putat{-85}{+115}{\colorbox{white}{$\Dsm\to\Km\Kp\pim$}}}
    \caption{The $M(\Dsm)$ distributions in the first $M(\pipipi)$ interval %
    for the tag modes of $\Dsm\to\KS\Km$ (top) and $\Dsm\to\Km\Kp\pim$ (bottom). Data (black points) are shown
    overlaid with the fit results including the total (solid blue),
signal PDF (dashed magenta), $\Dm$ background PDF (dotted red), and combinatorial background PDF (long-dashed blue) components.}
\label{fig:DTfitsvsm3pi_bin0}
\end{figure}

The number of $\Ds\to\pipipi X$ signal candidates in each $M(\pipipi)$ interval is determined by fitting the corresponding $M(\Dsm)$ distribution of the tag side.
From this fit, the raw signal yield $N^{\alpha}_{{\rm raw},i}$ is measured for each interval $i$ of the $M(\pipipi)$ distribution, for tag mode $\alpha$. 
The signal shape for each tag mode is
determined in
the corresponding ST fit (results shown in Fig.~\ref{fig:mDsfit}). The same shape is used in all $M(\pipipi)$ intervals, and the linear background shape parameter is left unconstrained in the fit.
The fitted raw signal yields ($N^{\alpha}_{{\rm raw},i}$) are presented in Tables~\ref{tab:yields_m400} and~\ref{tab:yields_m401} for the tag modes of $\Dsm\to \KS\Km$ and $\Dsm\to \Km\Kp\pim$, respectively. Two examples of the 22 fits are shown in Fig.~\ref{fig:DTfitsvsm3pi_bin0},
and the full set of fit results is made available as Supplemental Material~\cite{supp}.
Tables~\ref{tab:yields_m400} and~\ref{tab:yields_m401} summarize the 
observed signal yields,  
after subtracting the $\KS$-decay related and misID background contributions from the raw signal yields 
as described earlier in this section.

\begin{table*}[!htb]
\centering
    \caption{Raw signal yields and estimated background contributions for $D_s^-\to\KS\Km$ versus $\Ds\to\pipipi X$ events. The associated uncertainties are statistical only. The uncertainties from different background contributions are due to 
    the limited size of our inclusive MC simulation samples, whose integrated luminosity is 40 times of data. %
    }
\label{tab:yields_m400}
\begin{tabular}{l|cccccc}
\hline
    $M(\pipipi)$ interval & 1 & 2 &3 & 4 & 5 & 6 \\ \hline
Raw yield & 540.3 $\pm $ 28.7& 625.9 $\pm $ 31.8& 514.4 $\pm $ 28.4& 627.7 $\pm $ 31.3& 418.8 $\pm $ 26.1& 486.8 $\pm $ 27.2 \\ \hline
$\KS$ contribution & 16.6 $\pm $ 0.7& 37.9 $\pm $ 1.0& 28.3 $\pm $ 0.9& 22.4 $\pm $ 0.8& 12.3 $\pm $ 0.6& 11.3 $\pm $ 0.6\\ 
MisID contribution & 40.7 $\pm $ 1.1& 64.1 $\pm $ 1.4& 44.8 $\pm $ 1.1& 46.8 $\pm $ 1.2& 35.5 $\pm $ 1.0& 36.3 $\pm $ 1.0 \\ 
Total background & 57.4 $\pm $ 1.3& 102.0 $\pm $ 1.7\phantom{0}& 73.1 $\pm $ 1.5& 69.2 $\pm $ 1.4& 47.7 $\pm $ 1.2& 47.6 $\pm $ 1.2\\\hline
Background subtracted yield& 482.9 $\pm $ 28.8& 523.9 $\pm $ 31.8& 441.3 $\pm $ 28.4& 558.5 $\pm $ 31.3& 371.0 $\pm $ 26.1& 439.2 $\pm $ 27.2 \\ \hline
    $M(\pipipi)$ interval& 7& 8& 9& 10 & 11 & \\ \hline
Raw yield    &   403.3 $\pm $ 24.4& 515.3 $\pm $ 27.8& 400.5 $\pm $ 25.4& 494.8 $\pm $ 27.4& 153.3 $\pm $ 14.4 \\ \hline
$\KS$ contribution &  \phantom{0}6.8 $\pm $ 0.4& \phantom{0}9.7 $\pm $ 0.5& \phantom{0}8.1 $\pm $ 0.5& \phantom{0}2.7 $\pm $ 0.3& \phantom{0}0.4 $\pm $ 0.1 \\ 
MisID contribution &  28.5 $\pm $ 0.9& 35.9 $\pm $ 1.0& 36.5 $\pm $ 1.0& 27.6 $\pm $ 0.9& \phantom{0}0.8 $\pm $ 0.2 \\ 
Total background &  35.3 $\pm $ 1.0& 45.6 $\pm $ 1.1& 44.6 $\pm $ 1.1& 30.3 $\pm $ 0.9& \phantom{0}1.3 $\pm $ 0.2\\\hline
    Background subtracted yield & 368.1 $\pm $ 24.4& 469.6 $\pm $ 27.8& 356.0 $\pm $ 25.4& 464.5 $\pm $ 27.4& 152.0 $\pm $ 14.4 \\ \hline
\end{tabular}
\end{table*}

\begin{table*}[!htb]
\centering
    \caption{Raw signal yields and estimated background contributions for $D_s^-\to\Km\Kp\pim$ versus $\Ds\to\pipipi X$ events. The associated uncertainties are statistical only. The uncertainties from different background contributions are due to 
    the limited size of our inclusive MC simulation samples, whose integrated luminosity is 40 times of data. %
    }
\label{tab:yields_m401}
\begin{tabular}{l|cccccc}
\hline
    $M(\pipipi)$ interval& 1& 2& 3& 4& 5  & 6\\ \hline
Raw yield & 2300.8 $\pm $ 71.1& 2883.4 $\pm $ 85.8& 2405.7 $\pm $ 77.6& 2630.0 $\pm $ 81.3& 1944.3 $\pm $ 70.6& 2276.2 $\pm $ 74.3 \\ \hline    
$\KS$ contribution & \phantom{0}59.1 $\pm $ 1.3& 148.2 $\pm $ 2.1& 115.0 $\pm $ 1.8& \phantom{0}93.6 $\pm $ 1.7& \phantom{0}47.7 $\pm $ 1.2& \phantom{0}48.4 $\pm $ 1.2 \\ 
MisID contribution & 190.1 $\pm $ 2.4& 297.3 $\pm $ 3.0& 214.6 $\pm $ 2.5& 226.6 $\pm $ 2.6& 164.3 $\pm $ 2.2& 171.8 $\pm $ 2.2 \\ 
Total background & 249.2 $\pm $ 2.7& 445.5 $\pm $ 3.6& 329.6 $\pm $ 3.1& 320.2 $\pm $ 3.1& 212.0 $\pm $ 2.5& 220.2 $\pm $ 2.5\\\hline
Background subtracted yield& 2051.6 $\pm $ 71.1& 2437.9 $\pm $ 85.9& 2076.1 $\pm $ 77.7& 2309.8 $\pm $ 81.4& 1732.2 $\pm $ 70.7& 2056.0 $\pm $ 74.4 \\\hline
    $M(\pipipi)$ interval& 7& 8& 9& 10 & 11 &  \\ \hline
Raw yield & 1924.3 $\pm $ 65.2& 2182.5 $\pm $ 68.8& 1926.0 $\pm $ 65.5& 1993.0 $\pm $ 63.7& 767.7 $\pm $ 34.0 \\ \hline
$\KS$ contribution &  \phantom{0}31.1 $\pm $ 1.0& \phantom{0}36.9 $\pm $ 1.0& \phantom{0}31.1 $\pm $ 1.0& \phantom{0}10.7 $\pm $ 0.6& \phantom{0}1.4 $\pm $ 0.2 \\ 
MisID contribution &  127.5 $\pm $ 1.9& 169.0 $\pm $ 2.2& 168.3 $\pm $ 2.2& 126.3 $\pm $ 1.9& \phantom{0}3.9 $\pm $ 0.3 \\ 
Total background & 158.7 $\pm $ 2.2& 205.9 $\pm $ 2.5& 199.4 $\pm $ 2.4& 137.0 $\pm $ 2.0& \phantom{0}5.3 $\pm $ 0.4\\\hline
Background subtracted yield&  1765.6 $\pm $ 65.2& 1976.6 $\pm $ 68.8& 1726.6 $\pm $ 65.5& 1856.0 $\pm $ 63.7& 762.4 $\pm $ 34.0 \\ \hline
\hline
\end{tabular}
\end{table*}

Using the observed signal yields and the efficiency matrix elements
 $\epsilon^{\alpha}_{ij}$ presented in Tables~\ref{tab:effmat_m400} and~\ref{tab:effmat_m401}, the partial BFs 
in various $M(\pipipi)$ intervals are calculated using Eqs.~\ref{eq:solvmat} and~\ref{eq:pbf}
and shown in Table~\ref{tab:pbfsys} after combining the measurements from both
tag modes. 

\begin{table*}[!htb]
\centering
    \caption{Efficiency matrix $\epsilon_{ij}$ in percentage for $D_s^-\to\KS\Km$ versus $\Ds\to\pipipi X$ determined from signal MC simulated events. Each column gives the true $M(\pipipi)$ interval $j$, while each row gives the reconstructed $M(\pipipi)$ interval $i$. %
    }
\label{tab:effmat_m400}
\begin{tabular}{c|ccccccccccc}
\hline\hline
    $\epsilon_{ij}$
    & 1 & 2 &3 & 4 & 5 & 6 & 7 & 8 & 9 & 10 & 11\\ 
\hline
1 & 15.11  & \phantom{0}0.25  & \phantom{0}0.00  & \phantom{0}0.00  & \phantom{0}0.00  & \phantom{0}0.00  & \phantom{0}0.00  & \phantom{0}0.00  & \phantom{0}0.00  & \phantom{0}0.00  & \phantom{0}0.00 \\
2 & \phantom{0}0.19  & 15.86  & \phantom{0}0.67  & \phantom{0}0.08  & \phantom{0}0.01  & \phantom{0}0.00  & \phantom{0}0.00  & \phantom{0}0.00  & \phantom{0}0.00  & \phantom{0}0.00  & \phantom{0}0.00 \\
3 & \phantom{0}0.01  & \phantom{0}0.41  & 16.76  & \phantom{0}0.85  & \phantom{0}0.12  & \phantom{0}0.03  & \phantom{0}0.01  & \phantom{0}0.00  & \phantom{0}0.00  & \phantom{0}0.01  & \phantom{0}0.00 \\
4 & \phantom{0}0.01  & \phantom{0}0.02  & \phantom{0}0.62  & 20.99  & \phantom{0}1.00  & \phantom{0}0.19  & \phantom{0}0.05  & \phantom{0}0.01  & \phantom{0}0.00  & \phantom{0}0.00  & \phantom{0}0.00 \\
5 & \phantom{0}0.01  & \phantom{0}0.01  & \phantom{0}0.02  & \phantom{0}0.73  & 21.76  & \phantom{0}0.98  & \phantom{0}0.13  & \phantom{0}0.05  & \phantom{0}0.01  & \phantom{0}0.00  & \phantom{0}0.00 \\
6 & \phantom{0}0.00  & \phantom{0}0.00  & \phantom{0}0.01  & \phantom{0}0.03  & \phantom{0}0.88  & 21.93  & \phantom{0}1.18  & \phantom{0}0.21  & \phantom{0}0.05  & \phantom{0}0.00  & \phantom{0}0.00 \\
7 & \phantom{0}0.00  & \phantom{0}0.00  & \phantom{0}0.00  & \phantom{0}0.01  & \phantom{0}0.04  & \phantom{0}0.90  & 22.46  & \phantom{0}0.91  & \phantom{0}0.11  & \phantom{0}0.00  & \phantom{0}0.00 \\
8 & \phantom{0}0.00  & \phantom{0}0.00  & \phantom{0}0.00  & \phantom{0}0.00  & \phantom{0}0.01  & \phantom{0}0.06  & \phantom{0}1.11  & 23.46  & \phantom{0}0.88  & \phantom{0}0.05  & \phantom{0}0.00 \\
9 & \phantom{0}0.00  & \phantom{0}0.00  & \phantom{0}0.00  & \phantom{0}0.00  & \phantom{0}0.00  & \phantom{0}0.01  & \phantom{0}0.03  & \phantom{0}0.66  & 24.84  & \phantom{0}0.77  & \phantom{0}0.02 \\
10 & \phantom{0}0.00  & \phantom{0}0.00  & \phantom{0}0.00  & \phantom{0}0.00  & \phantom{0}0.00  & \phantom{0}0.00  & \phantom{0}0.00  & \phantom{0}0.01  & \phantom{0}0.48  & 27.10  & \phantom{0}0.75 \\
11 & \phantom{0}0.00  & \phantom{0}0.00  & \phantom{0}0.00  & \phantom{0}0.00  & \phantom{0}0.00  & \phantom{0}0.00  & \phantom{0}0.00  & \phantom{0}0.00  & \phantom{0}0.00  & \phantom{0}0.10  & 24.53 \\
\hline\hline
\end{tabular}
\end{table*}

\begin{table*}[!htb]
\centering
    \caption{Efficiency matrix $\epsilon_{ij}$ in percentage for $D_s^-\to\Km\Kp\pim$ versus $\Ds\to\pipipi X$ determined from signal MC simulated events. Each column gives the true $M(\pipipi)$ interval $j$, while each row gives the reconstructed $M(\pipipi)$ interval $i$. %
    }
\label{tab:effmat_m401}
\begin{tabular}{c|ccccccccccc}
\hline
\hline
$\epsilon_{ij}$ 
    & 1 & 2 &3 & 4 & 5 & 6 & 7 & 8 & 9 & 10 & 11\\ \hline
1 & 12.80  & \phantom{0}0.20  & \phantom{0}0.00  & \phantom{0}0.00  & \phantom{0}0.00  & \phantom{0}0.00  & \phantom{0}0.00  & \phantom{0}0.00  & \phantom{0}0.00  & \phantom{0}0.00  & \phantom{0}0.00 \\
2 & \phantom{0}0.17  & 13.65  & \phantom{0}0.54  & \phantom{0}0.06  & \phantom{0}0.01  & \phantom{0}0.00  & \phantom{0}0.00  & \phantom{0}0.00  & \phantom{0}0.00  & \phantom{0}0.00  & \phantom{0}0.00 \\
3 & \phantom{0}0.01  & \phantom{0}0.33  & 14.64  & \phantom{0}0.67  & \phantom{0}0.09  & \phantom{0}0.02  & \phantom{0}0.00  & \phantom{0}0.00  & \phantom{0}0.00  & \phantom{0}0.00  & \phantom{0}0.00 \\
4 & \phantom{0}0.00  & \phantom{0}0.01  & \phantom{0}0.47  & 17.96  & \phantom{0}0.96  & \phantom{0}0.14  & \phantom{0}0.04  & \phantom{0}0.01  & \phantom{0}0.00  & \phantom{0}0.00  & \phantom{0}0.00 \\
5 & \phantom{0}0.00  & \phantom{0}0.00  & \phantom{0}0.01  & \phantom{0}0.62  & 18.36  & \phantom{0}0.83  & \phantom{0}0.13  & \phantom{0}0.05  & \phantom{0}0.00  & \phantom{0}0.00  & \phantom{0}0.00 \\
6 & \phantom{0}0.00  & \phantom{0}0.00  & \phantom{0}0.01  & \phantom{0}0.03  & \phantom{0}0.83  & 18.89  & \phantom{0}1.01  & \phantom{0}0.16  & \phantom{0}0.04  & \phantom{0}0.00  & \phantom{0}0.00 \\
7 & \phantom{0}0.00  & \phantom{0}0.00  & \phantom{0}0.00  & \phantom{0}0.01  & \phantom{0}0.03  & \phantom{0}0.75  & 19.19  & \phantom{0}0.77  & \phantom{0}0.09  & \phantom{0}0.01  & \phantom{0}0.00 \\
8 & \phantom{0}0.00  & \phantom{0}0.00  & \phantom{0}0.00  & \phantom{0}0.00  & \phantom{0}0.01  & \phantom{0}0.06  & \phantom{0}0.88  & 20.16  & \phantom{0}0.76  & \phantom{0}0.06  & \phantom{0}0.00 \\
9 & \phantom{0}0.00  & \phantom{0}0.00  & \phantom{0}0.00  & \phantom{0}0.00  & \phantom{0}0.00  & \phantom{0}0.01  & \phantom{0}0.02  & \phantom{0}0.56  & 21.15  & \phantom{0}0.55  & \phantom{0}0.01 \\
10 & \phantom{0}0.00  & \phantom{0}0.00  & \phantom{0}0.00  & \phantom{0}0.00  & \phantom{0}0.00  & \phantom{0}0.00  & \phantom{0}0.00  & \phantom{0}0.01  & \phantom{0}0.40  & 23.13  & \phantom{0}0.67 \\
11 & \phantom{0}0.00  & \phantom{0}0.00  & \phantom{0}0.00  & \phantom{0}0.00  & \phantom{0}0.00  & \phantom{0}0.00  & \phantom{0}0.00  & \phantom{0}0.00  & \phantom{0}0.00  & \phantom{0}0.07  & 21.29 \\
\hline
\hline
\end{tabular}
\end{table*}

\begin{table}[!htb]
\centering
    \caption{Partial BFs (in percentage) of the decay $\Ds\to\pipipi X$ combined from the two tag modes. The first and second uncertainties are statistical and systematic, respectively. }
\label{tab:pbfsys}
\begin{tabular}{cc}
\hline\hline
    $M(\pipipi)$ interval & ${\Delta {\cal B}}_{3\pi X,i}$ (\%)  \\  \hline
1 & $4.63 \pm 0.14 \pm 0.14$ \\
2    & $4.92 \pm 0.16 \pm 0.19$ \\ 
 3   & $3.79 \pm 0.13 \pm 0.10$ \\
 4   & $3.55 \pm 0.12 \pm 0.09$  \\ 
5 & $2.42 \pm 0.10 \pm 0.07$ \\
6    & $2.87 \pm 0.10 \pm 0.09$ \\
7    & $2.39 \pm 0.09 \pm 0.07$ \\
8    & $2.69 \pm 0.09 \pm 0.07$  \\ 
9 & $2.19 \pm 0.08 \pm 0.05$ \\
10    & $2.32 \pm 0.07 \pm 0.05$ \\
11    & $1.01 \pm 0.04 \pm 0.04$  \\ 
\hline\hline
\end{tabular}
\end{table}

\section {Systematic uncertainties}
\label{sec:sys}

The methods to determine the systematic
uncertainties in the measured total and partial BFs
of inclusive $\Ds\to\pipipi X$ decays %
 are described below.

\subsection{\boldmath Signal shape in $M(\Dsm)$}
The signal $M(\Dsm)$ PDF is changed from the DSCB shape to a shape modeled with signal MC simulated events convolved with a Gaussian function. 
        The width and mean parameters of the Gaussian function are floating in the fits to account for the resolution difference between data and simulation. Alternatively, the fits
        are performed by allowing the two signal shape parameters $m_0$ and $\sigma_0$ in different $M(\pipipi)$ intervals to differ. %
        The largest variations are taken as the relative systematic uncertainties.

\subsection{\boldmath Background shape in $M(\Dsm)$ }
The background shape is changed from the linear PDF to an exponential or a histogram PDF based on MC simulation. Also  the fits are performed with the yields of $\Dm\to\KS\pim$ ($\Dm\to\Kp\pim\pim$) in the tag mode $\Dsm\to\KS\Km$ ($\Dsm\to\Km\Kp\pim$) floating, rather than fixed to the numbers from simulation.
        The largest variations are taken as the relative systematic uncertainties.

        \subsection{ Pion %
        tracking and PID efficiency} %
The $\pi^{\pm}$ tracking efficiency is studied with the control sample of
$\epem\to K^+ K^-\pip\pim$ decay. The data-MC tracking efficiency ratios of $\pip$ and $\pim$ are measured in bins of pion transverse momentum. The signal MC samples are then weighted by these ratios to recalculate signal efficiency matrices. This leads to 0.98\% variation in total BF, which is taken as the systematic uncertainty associated with tracking efficiency.

The $\pi^{\pm}$ PID efficiency is studied with the control samples of
$\epem\to K^+ K^-\pip\pim(\piz)$ and $\pip\pim\pip\pim(\piz)$ decays. The data-MC tracking efficiency ratios of $\pi^{\pm}$ are measured in bins of pion momentum. The signal MC samples are then weighted by these ratios to recalculate signal efficiency matrices. This leads to 0.99\% variation in the total BF, which is taken as the systematic uncertainty associated with PID efficiency.
%        Based on the same methods detailed in Ref.~\cite{BESIIIkkpipi0}, the uncertainty associated with the efficiency of $\pi^{\pm}$ tracking or PID per pion is 0.5\%. 
%        Therefore, the total systematic uncertainty on either tracking or PID efficiency of the $3\pi$ system of the signal side is 1.5\%.

        \subsection{\boldmath Background from $\KS$ decays}
        As $N^{\alpha}_{{\KS},i}$ defined in Eq.~(\ref{eq:raw2obs}) is determined by the $D_s$ inclusive MC simulation sample, the data and simulation differences on the $\Ds\to\KS\pip X$ contributions are studied by using control samples with dedicated stringent requirements on the $3\pi$ system: one $\pipi$ pair forms the reconstructed $\KS$ that has $L/\sigma_L>2$ %
        and $|M(\pipi)-m_{\KS}|<10 $\mevcc. These requirements lead to
        a data sample with more than 99\% of the $\Ds$ candidates having a $\pipi$ pair from $\KS$.
        The agreement on the observed number of $\Ds$ candidates in data and the number from simulation is at $\sim$10\% level across different $M(\pipipi)$ intervals. 
        Therefore, the $N^{\alpha}_{{\KS},i}$ values are all scaled by $\pm 10$\% in turn, and the BFs 
        are recalculated according to altered $N^{\alpha}_{{\rm obs},i}$ values based on Eq.~(\ref{eq:raw2obs}).
        The larger variations from
        the two measurements are taken as the relative systematic uncertainties on the $\KS$ background. 
        
        Furthermore, the systematic uncertainty related to the $\KS$ veto of $|M(\pipi)-m_{\KS}|<10$~\mevcc\ 
        is examined by varying the mass window by $\pm 5$~\mevcc. The resulting changes in the BFs        
        are smaller than the statistical uncertainties on the differences, so no uncertainty is
        assigned for this source according to Ref.~\cite{barlow}.

\subsection{Particle misidentification background}
        The $D_s$ inclusive MC simulation sample is also used to estimate $N^{\alpha}_{{\rm misID},i}$ defined in Eq.~(\ref{eq:raw2obs}). %
        The uncertainties on $N^{\alpha}_{{\rm misID},i}$ come from limited knowledge on the BFs of different decays that contribute to the misID backgrounds,
        and differences in misID rates between data and simulation, that are
        considered separately in the three misID scenarios as listed in Sec.~\ref{sec:pidbkg}.

        The $\Ds\to\pipi\ell^+ X$ decays form the background where the leptons $\ell=e, \mu$ are misidentified as pions, and the decays $\Ds\to \Km\pip\pip X$ and $\Ds\to \Kp\pipi X$
        are relevant for the background when a kaon is misidentified instead. The $\Ds\to\pipi\ell^+ X$ decays
        are dominated by the semileptonic decays of $\Ds\to\phi\ell^+ \nu_\ell$ and $\Ds\to\eta^{(\prime)}\ell^+ \nu_\ell$. By assuming LFU in semileptonic charm decays, 
        the data and simulation differences on modeling the $\Ds\to\pipi\ell^+ X$ decays are studied using samples of 
        $\Ds\to\pipi\ep X$. 
        This LFU assumption has been found to hold experimentally, {\it e.g.} within a precision of 1.4\% in $D^0\to K^- \ell^+\nu_{\ell}$ decays~\cite{bes3d2klnu}.
        The electron PID requirement used on one of the positively charged track in the  3$\pi$ system is the same as that in Ref.~\cite{BESIIIDs2eX}. 
        As the relative difference between data and MC simulation is below 5\%, the sums of $N^{\alpha}_{e\to\pi,\ i}+N^{\alpha}_{\mu\to\pi,\ i}$ are
        all scaled
        by $\pm$5\% in turn. The larger variations from the two measurements are taken as the systematic uncertainties related to the lepton misID 
        background. Similarly, the dedicated simulation samples of $\Ds\to \Km\pip\pip X$ and $\Ds\to \Kp\pipi X$ decays can be compared to data by
        imposing kaon identification requirements on one track from the 3$\pi$ system. This study
        indicates a disagreement between data and simulation at a level of about 15\%. This is significantly larger than in the lepton
        case due to the poor knowledge of hadronic $\Ds$ decays with at least one charged kaon.
        The $N^{\alpha}_{K\to\pi,\ i}$ values are all scaled by $\pm$15\% in turn, and the larger variations from the two measurements are taken as the systematic uncertainties related to the kaon misID. 
        Separate studies %
        on the probabilities of tracks being misidentified as pions 
        show the effects due to data and simulation differences are negligible. Therefore, this is not included as part of the systematic uncertainties. 

        \subsection{Pion multiplicity}
        The number of charged pions in $\Ds\to \pipipi X$ decays
         may be inaccurately modeled in the simulation mainly due to the limited knowledge of hadronic $\Ds$ decays
         with five or more charged pions in the final states such as $\Ds\to \pipipi\pipi \piz(\piz)$~\cite{pdg}.          
         The signal MC simulated events are weighted so that the simulated distribution of the number of 
         charged pion candidates at the signal side matches with that in the data. 
         This causes changes in signal efficiencies, and the resulting BF variations 
         using the weighted MC simulated events are taken as the relative systematic uncertainties.

        \subsection{Pion momenta}%
        \label{sec:Ppi}
         The signal MC simulated events are weighted to improve the agreement of the pion momentum distributions between data and simulation. 
         The multivariate algorithm~\cite{gbrw} to determine the weights is 
trained on the distributions from unweighted signal MC simulation
and background subtracted data samples as shown in Fig.~\ref{fig:Ppi3}. 
         Using the weighted MC candidates based on the three-dimensional weighting scheme
         to recalculate signal efficiency matrices for both tag modes, 
        the variations in the BF central values are taken as the relative systematic uncertainties.

        As the data-MC differences in the angular distributions of pion momenta may result
        in differences in the mass spectra of the 3$\pi$ system, 
        we also adopt a five-dimensional weighting scheme with the addition of $M_{\rm high}(\pipi)$ and $M_{\rm low}(\pipi)$
        to account for the data-MC differences shown in Fig.~\ref{fig:M3pi}.
        The resulting variations in BFs are smaller than those from the three-dimensional weighting scheme, and thus no additional
        uncertainty is assigned.

        \subsection{Monte Carlo simulation sample size}
        \label{sec:mcstat}
        The BFs are recalculated 1000 
        times by randomly perturbing the efficiency matrix elements shown in Tables~\ref{tab:effmat_m400} and~\ref{tab:effmat_m401} according to their uncertainties. The distributions of the recalculated BF central values are fitted with a Gaussian function and the fitted Gaussian widths are taken as the relative systematic uncertainties. 

    \subsection{Measurement bias} 
     \label{sec:SCS}
    The uncertainties due to the entire analysis procedure to extract the BFs
        are estimated by using 40 inclusive MC simulation samples, each with a total candidate number statistically
        matched to that in the data. While good agreement between the input and measured BF values is found, 
        the mean biases are taken as the relative systematic uncertainties.

     \subsection{Signal MC simulation model}
     The purpose of measuring ${\cal B}(\Ds\to\pipipi X)$ in intervals of 
     $M(\pipipi)$ is to be independent of the $\Ds$ decay model 
     used in the signal simulation. 
     A number of exclusive $\Ds$ decay modes such as $\Ds\to\etapr\rho^+$, $\Ds\to\phi\pipipi$, and $\Ds\to \phi\rho^+$ are used to perform the tests. The yield of the respective signal MC simulation events is weighted by about 40\%. This in effect changes the related input BF by about 40\% (up or down). 
     The signal efficiency matrices are then updated based on the weighted MC simulation samples. The obtained BF results show good agreement with the baseline results well within one standard deviation. Therefore, a systematic uncertainty related to the signal MC simulation model is not assigned. 

     \subsection{Single-candidate requirement}
     \label{sec:SCS}
     The single-candidate requirement described in Sec.~\ref{sec:st} may bias the measurement due to potentially different candidate multiplicities and $M_{\rm rec}$ distributions in data and MC simulation. The systematic uncertainties from the single-candidate requirement are examined by rerunning the measurement with this requirement removed. 
     The resulting BF variations are smaller than the statistical uncertainties on the differences, so no uncertainty is assigned for this source.

\subsection{Summary of systematic uncertainties}        
Table~\ref{tab:finalsys} summarizes
contributions from the different systematic sources in the measurement of ${\cal B}(\Ds\to\pipipi X)$. The largest sources of systematic uncertainties 
come from tracking and PID of the $3\pi$ system.  These
contributions are combined in quadrature to determine the total
systematic uncertainty in the BF measurement. The summary table for the systematic uncertainties of 
the partial BFs of $\Ds\to\pipipi X$ is given in the Supplemental Material~\cite{supp}.

\begin{table}[!htb]
\centering
    \caption{Sources of systematic uncertainties in the measurement of ${\cal B}(\Ds\to\pipipi X)$. %
    }
\label{tab:finalsys}
\begin{tabular}{lc}
\hline    
\hline
    Source & Relative uncertainty  (\%) \\\hline
    Signal shape  & 0.49\\
Background shape &  0.40 \\
%
%    $\pipipi$ tracking efficiency & 1.50  \\
%$\pipipi$ PID efficiency & 1.50 \\
    $\pipipi$ tracking efficiency & 0.98  \\
$\pipipi$ PID efficiency & 0.99 \\
    $\KS$ background & 0.34  \\
    Kaon misID background & 0.32  \\
    Lepton misID background & 0.35 \\
    Pion multiplicity  & 0.51 \\
    Pion momenta  & 0.86 \\
    MC simulation sample size & 0.01 \\
    Fit bias  & 0.03  \\
\hline
%
%
%Total & 2.50 \\
Total & 1.92 \\
\hline\hline
\end{tabular}
\end{table}
         
\section {Summary}
\label{sec:CONLUSION}
\begin{figure}[!htb]
\centering
\includegraphics[width=0.95\linewidth]{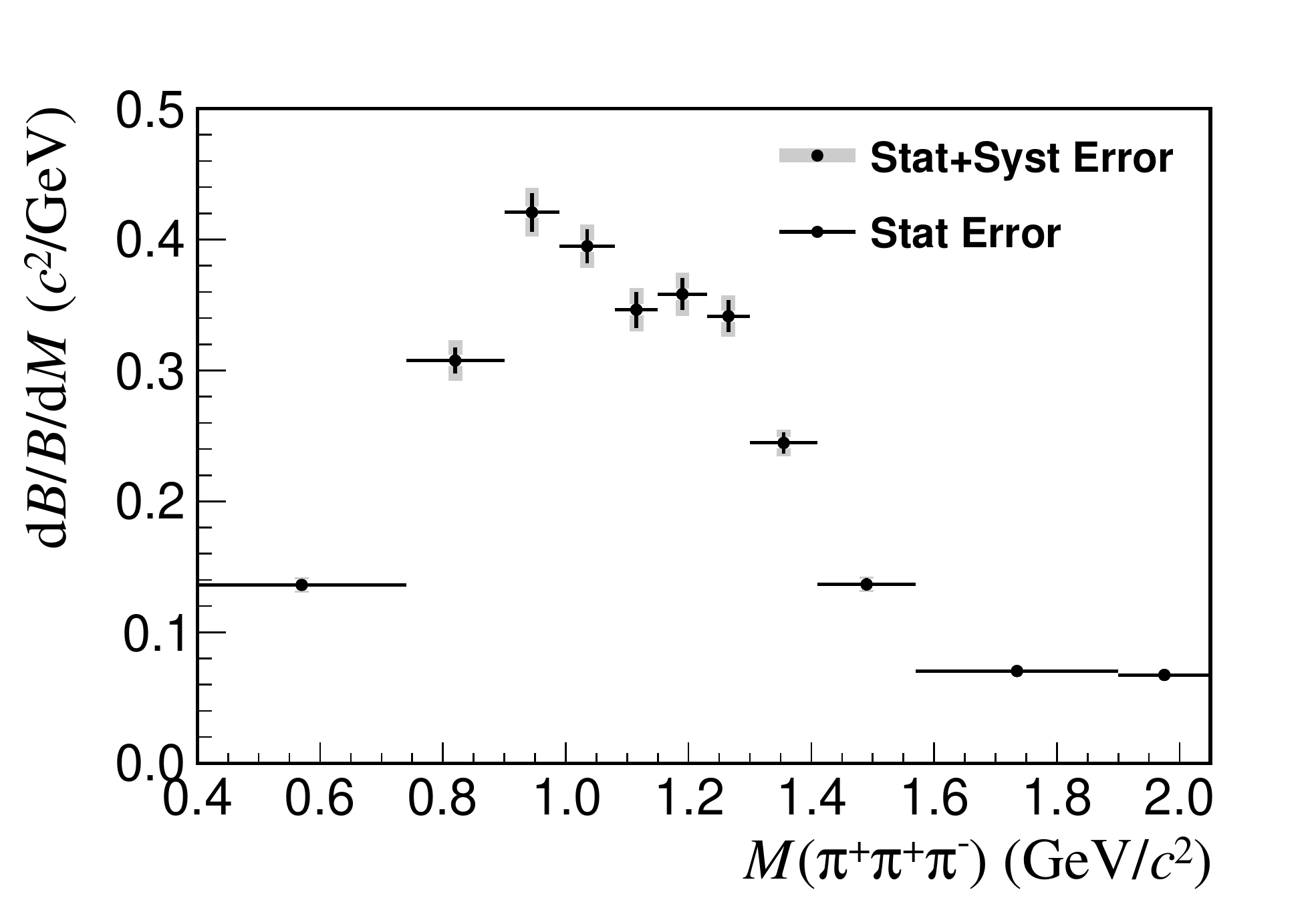}
    \caption{Partial BFs normalized by the total BF and interval widths as a function of $M(\pipipi)$ for the $\Ds\to\pipipi X$ decay. 
%    The vertical uncertainties shown on the points are the quadratic sums of the statistical and systematic uncertainties.
    The vertical gray lines represent the uncertainties as the quadratic sums of the statistical and systematic uncertainties.
    }
\label{fig:pbf_3piX}
\end{figure}
Based on 3.19~\invfb of $\epem$ collision data recorded with the BESIII detector at $E_{\rm cm} = 4.178$~\gev, 
the BF of the inclusive $\Ds\to\pipipi X$ decay is measured for the first time. It is found
to be
\[{\cal B}(\Ds\to\pipipi X) = \left(32.81\pm0.35_{\rm stat}\pm {0.63}_{\rm syst}\right)\%.\]
This is larger than the sum of all observed exclusive BFs of $\sim$25\% based on the PDG and recent measurements as summarized in Table~\ref{tab:exclmodes}, hinting at potentially unobserved decay modes with at least three charged pions in the final states.

Furthermore, the partial BFs of %
the $\Ds\to\pipipi X$ decay as a function of $M(\pipipi)$ are also measured, as presented in Table~\ref{tab:pbfsys} and Fig.~\ref{fig:pbf_3piX}. 
As the feed-up contribution from lower intervals is expected to be negligible ($\sim$1\% of exclusive signals), the partial BF in the last $M(\pipipi)$ interval is consistent with the 
exclusive BF of ${\cal B}(\Ds\to\pipipi) = (1.08\pm0.04)$\% from the PDG~\cite{pdg}, with 
somewhat lower precision. The measured total and partial BFs of $\Ds\to\pipipi X$ offer important inputs to constrain the systematic uncertainties in future LHCb measurements on $R(D^{*-})$ and $R(\Lambda_c^+)$ with much larger data 
samples~\cite{lhcb-white-paper}.

\section*{\boldmath ACKNOWLEDGMENTS}
The BESIII collaboration thanks the staff of BEPCII and the IHEP computing center for their strong support. This work is supported in part by National Key R\&D Program of China under Contracts Nos. 2020YFA0406400, 2020YFA0406300; National Natural Science Foundation of China (NSFC) under Contracts Nos. 11635010, 11735014, 11835012, 11935015, 11935016, 11935018, 11961141012, 12022510, 12025502, 12035009, 12035013, 12192260, 12192261, 12192262, 12192263, 12192264, 12192265; the Chinese Academy of Sciences (CAS) Large-Scale Scientific Facility Program; Joint Large-Scale Scientific Facility Funds of the NSFC and CAS under Contract No. U1832207, U1932108; the CAS Center for Excellence in Particle Physics
(CCEPP); 100 Talents Program of CAS; The Institute of Nuclear and Particle Physics (INPAC) and Shanghai Key Laboratory for Particle Physics and Cosmology; ERC under Contract No. 758462; European Union's Horizon 2020 research and innovation programme under Marie Sklodowska-Curie grant agreement under Contract No. 894790; German Research Foundation DFG under Contracts Nos. 443159800, 455635585, Collaborative Research Center CRC 1044, FOR5327, GRK 2149; Istituto Nazionale di Fisica Nucleare, Italy; Ministry of Development of Turkey under Contract No. DPT2006K-120470; National Science and Technology fund; National Science Research and Innovation Fund (NSRF) via the Program Management Unit for Human Resources \& Institutional Development, Research and Innovation under Contract No. B16F640076; Olle Engkvist Foundation under Contract No. 200-0605; STFC (United Kingdom); Suranaree University of Technology (SUT), Thailand Science Research and Innovation (TSRI), and National Science Research and Innovation Fund (NSRF) under Contract No. 160355; The Royal Society, UK under Contracts Nos. DH140054, DH160214; The Swedish Research Council; U. S. Department of Energy under Contract No. DE-FG02-05ER41374.

\end{document}

\newpage
\clearpage

\title{Supplementary material}
\date{\vspace{-5ex}}

\maketitle

In addition to the results presented in the main body of the paper,
we provide a supplementary collection of figures on the fits used in determining the double-tag yields in the two tag modes, and a table summarizing the systematic uncertainties of the partial BFs of $D_s^+\to\pipipi X$ in intervals of $M(\pipipi)$. 

\begin{figure}[!htb]
\centering
\includegraphics[width=0.48\linewidth]{datafit_m3pibin_0_m400.pdf}
\includegraphics[width=0.48\linewidth]{datafit_m3pibin_0_m401.pdf}\\
\includegraphics[width=0.48\linewidth]{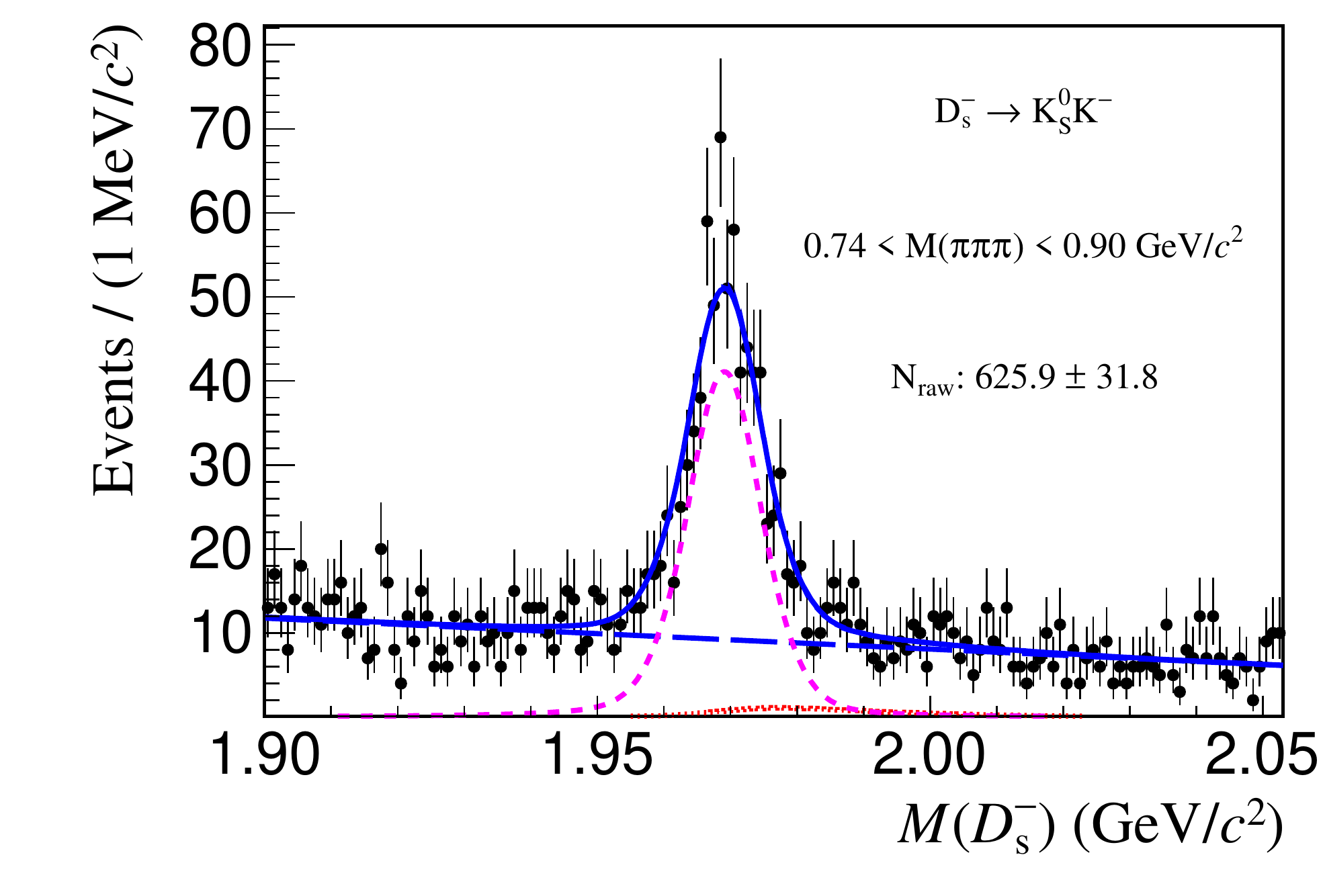}
\includegraphics[width=0.48\linewidth]{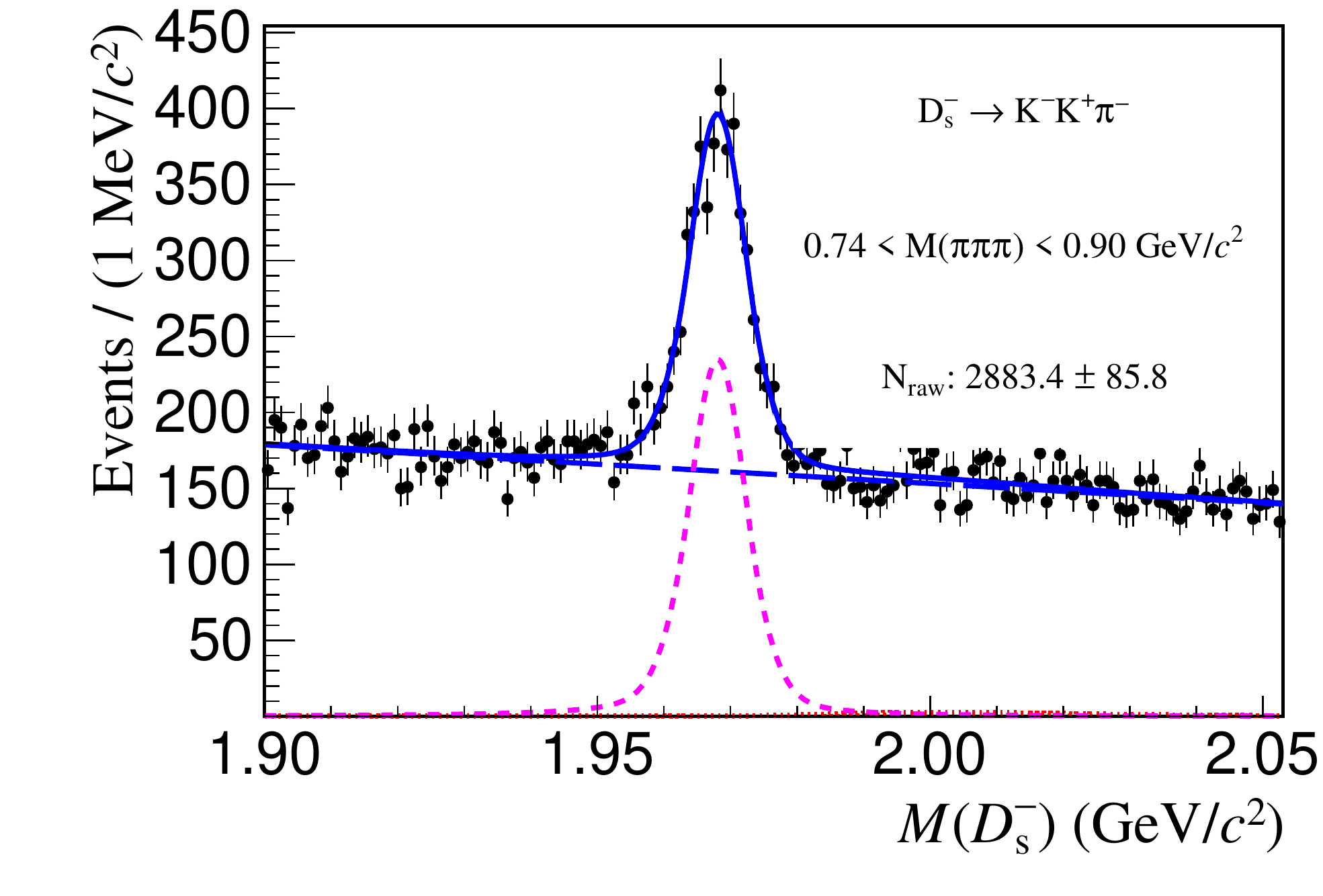}\\
\includegraphics[width=0.48\linewidth]{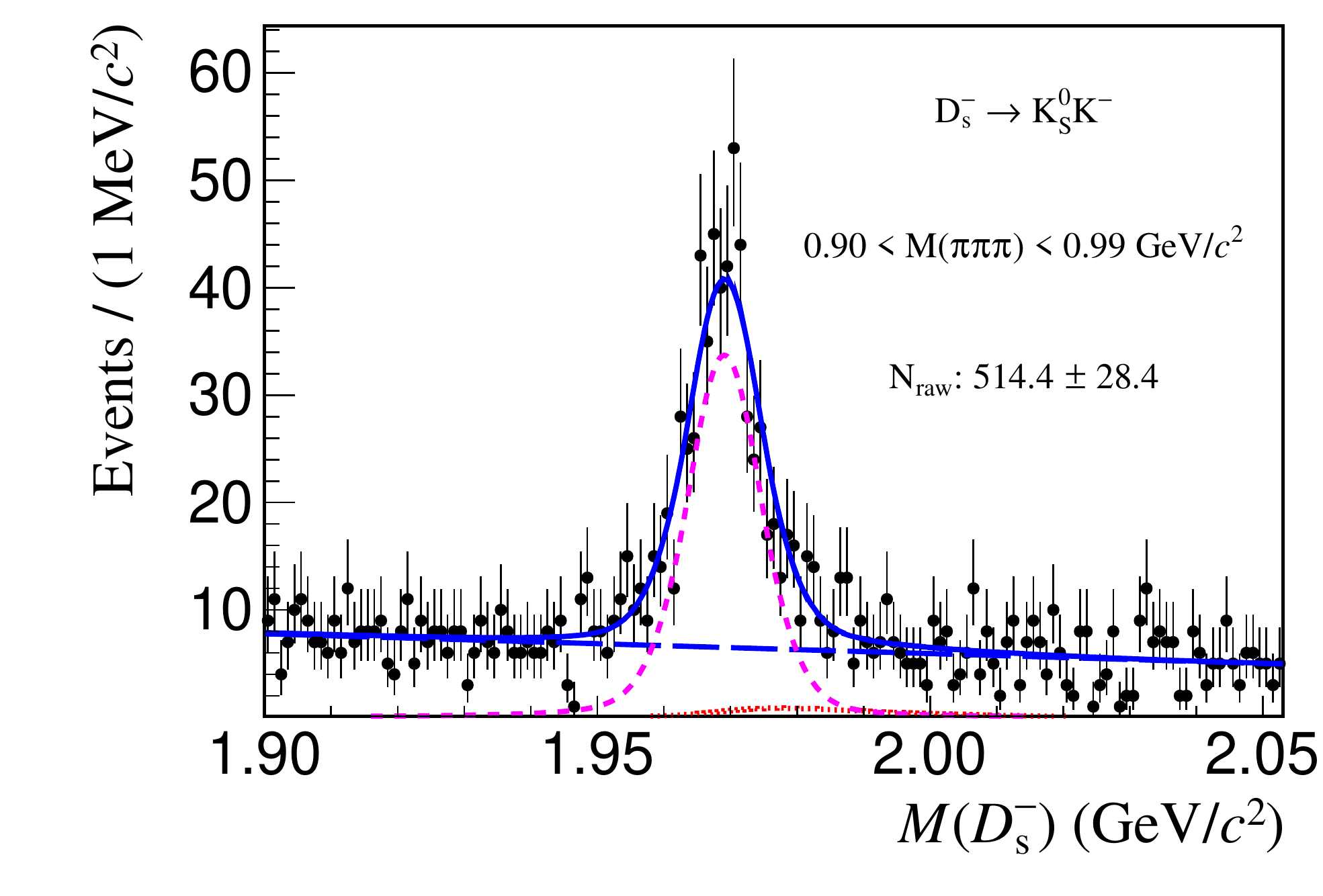}
\includegraphics[width=0.48\linewidth]{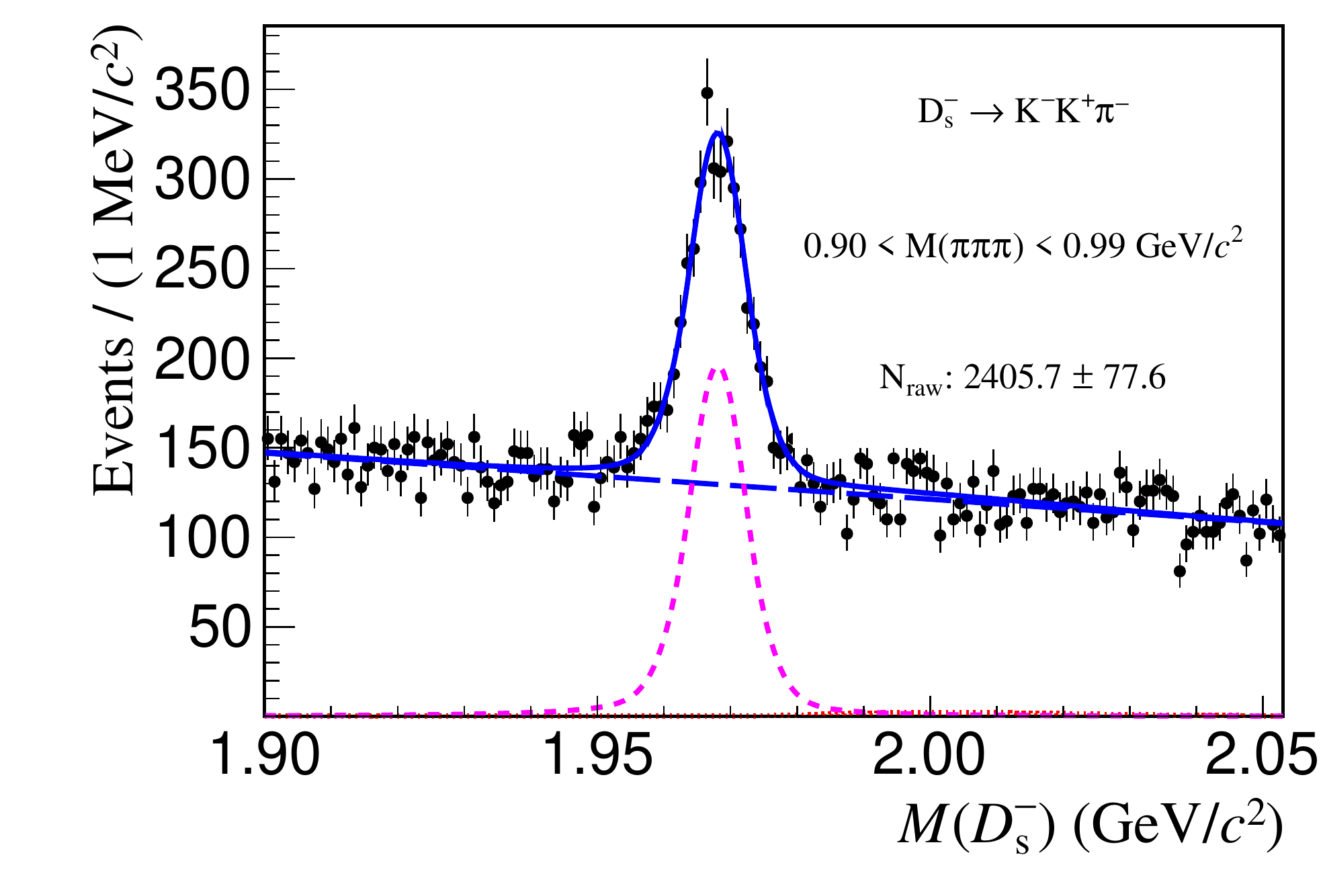}\\
\includegraphics[width=0.48\linewidth]{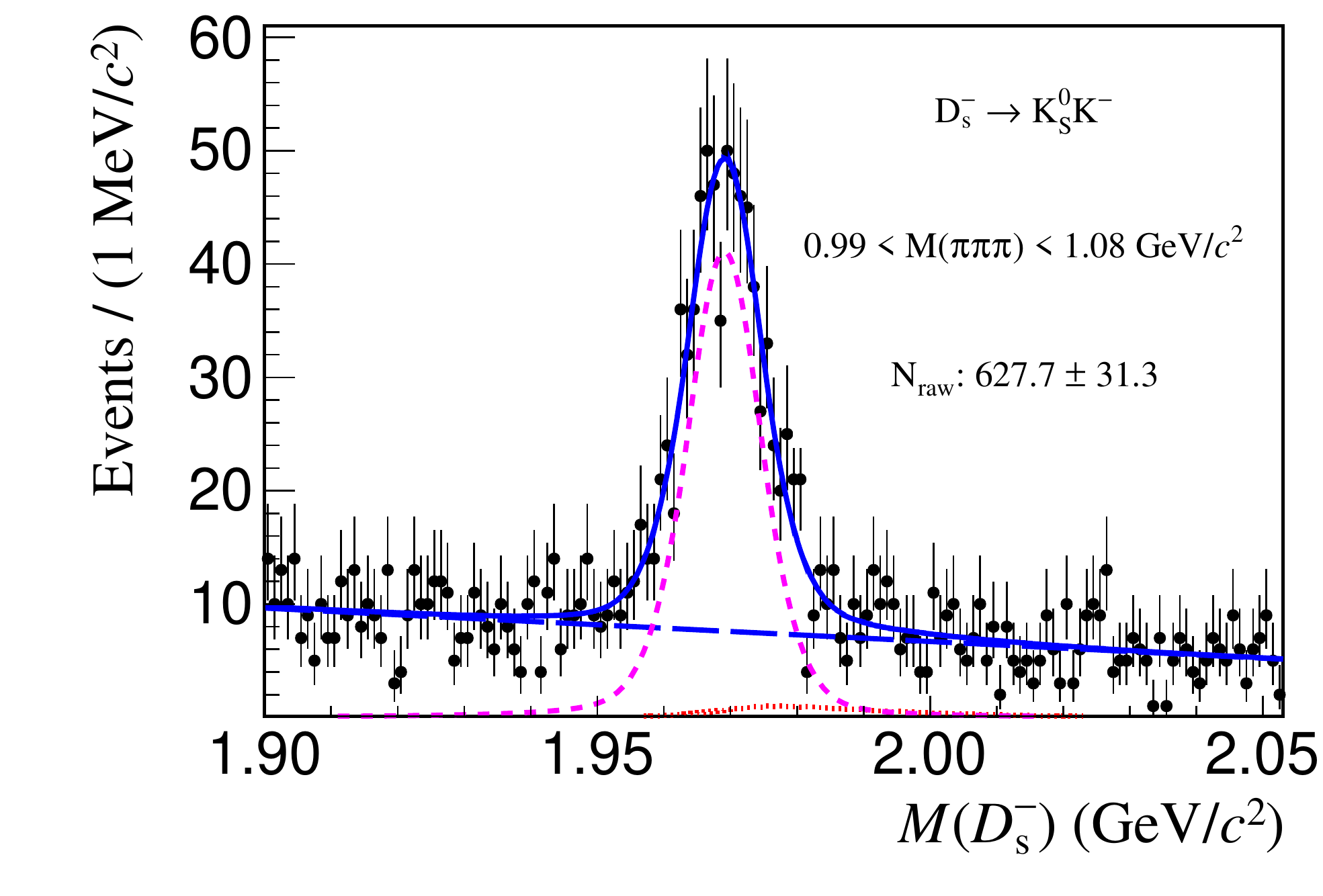}
\includegraphics[width=0.48\linewidth]{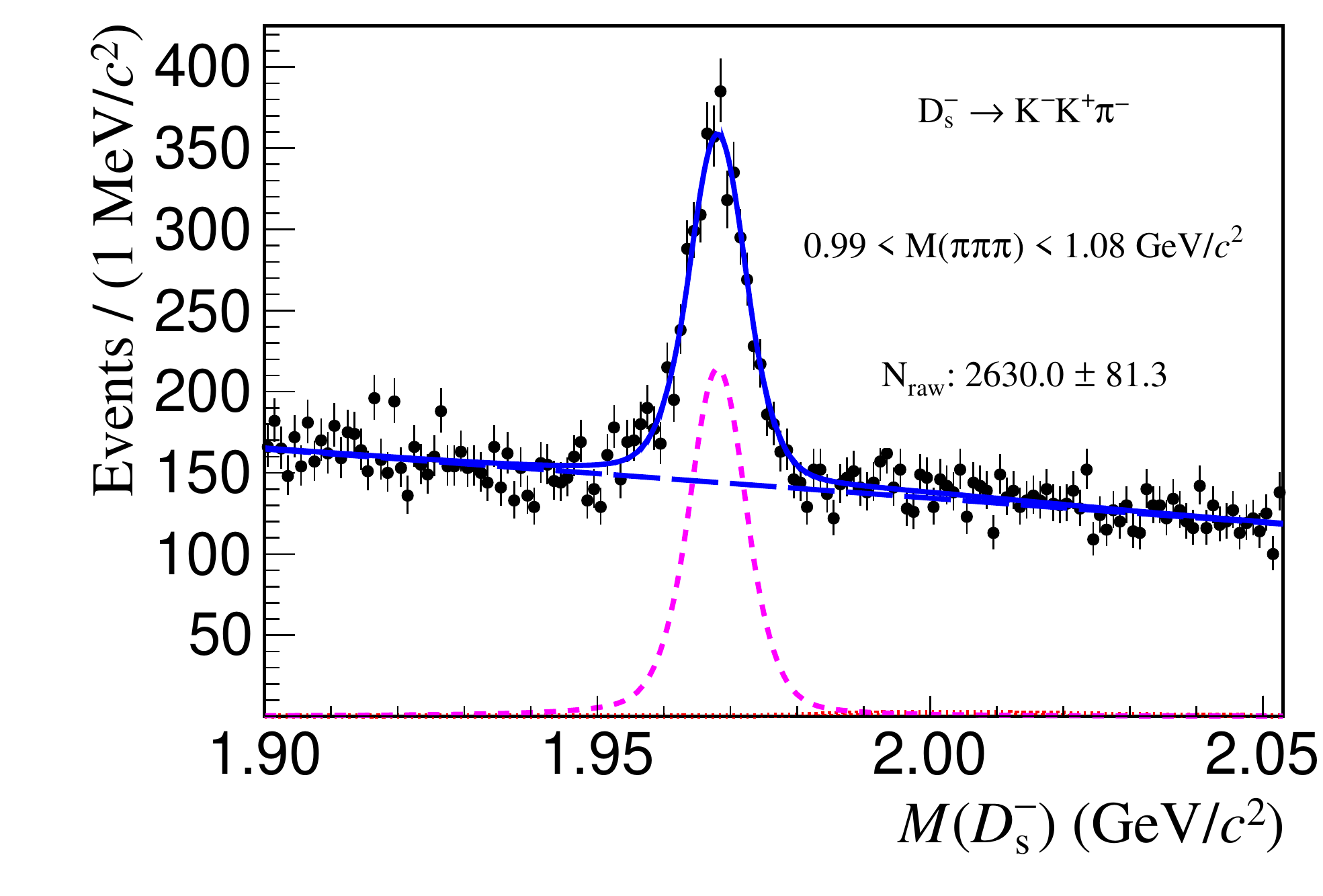}
%\begin{quote}
    \caption{The invariant mass $M(\Dsm)$ distributions %of the ST signals ($N^{\alpha}_{\rm ST}$) of 
    for the ST modes $\Dsm\to K_S^0K^-$ (left) and $\Dsm\to K^- K^+ \pi^-$ (right), part 1/3. Data (black points) are shown
    overlaid with the fit results including the total (solid blue),
signal PDF (dashed magenta), $\Dm$ background PDF (dotted red), and combinatorial background PDF (long-dashed blue) components.}
\label{fig:DTfitsvsm3pi_p1}
%\end{quote}
\end{figure}
\begin{figure}[!htb]
\centering
\includegraphics[width=0.48\linewidth]{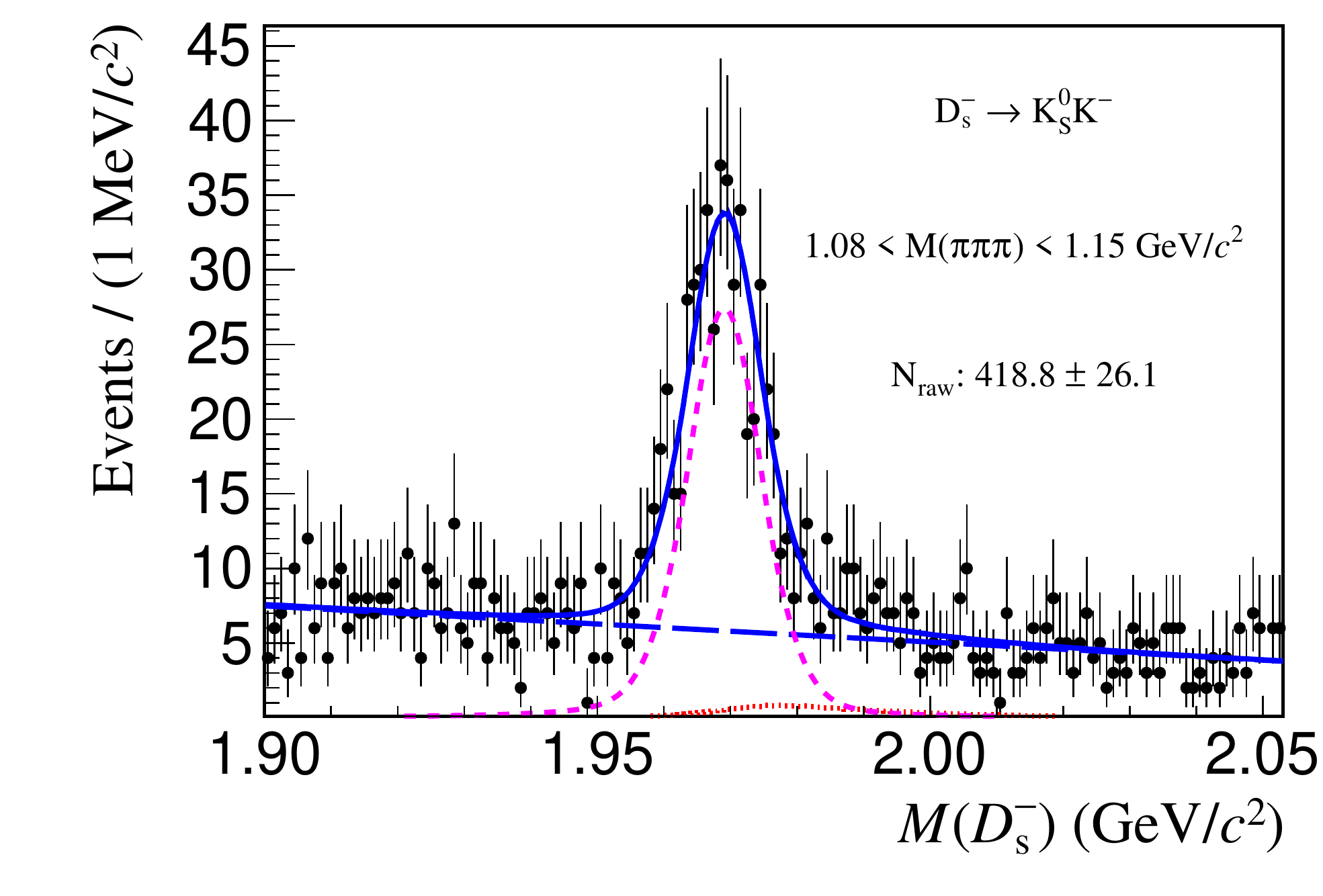}
\includegraphics[width=0.48\linewidth]{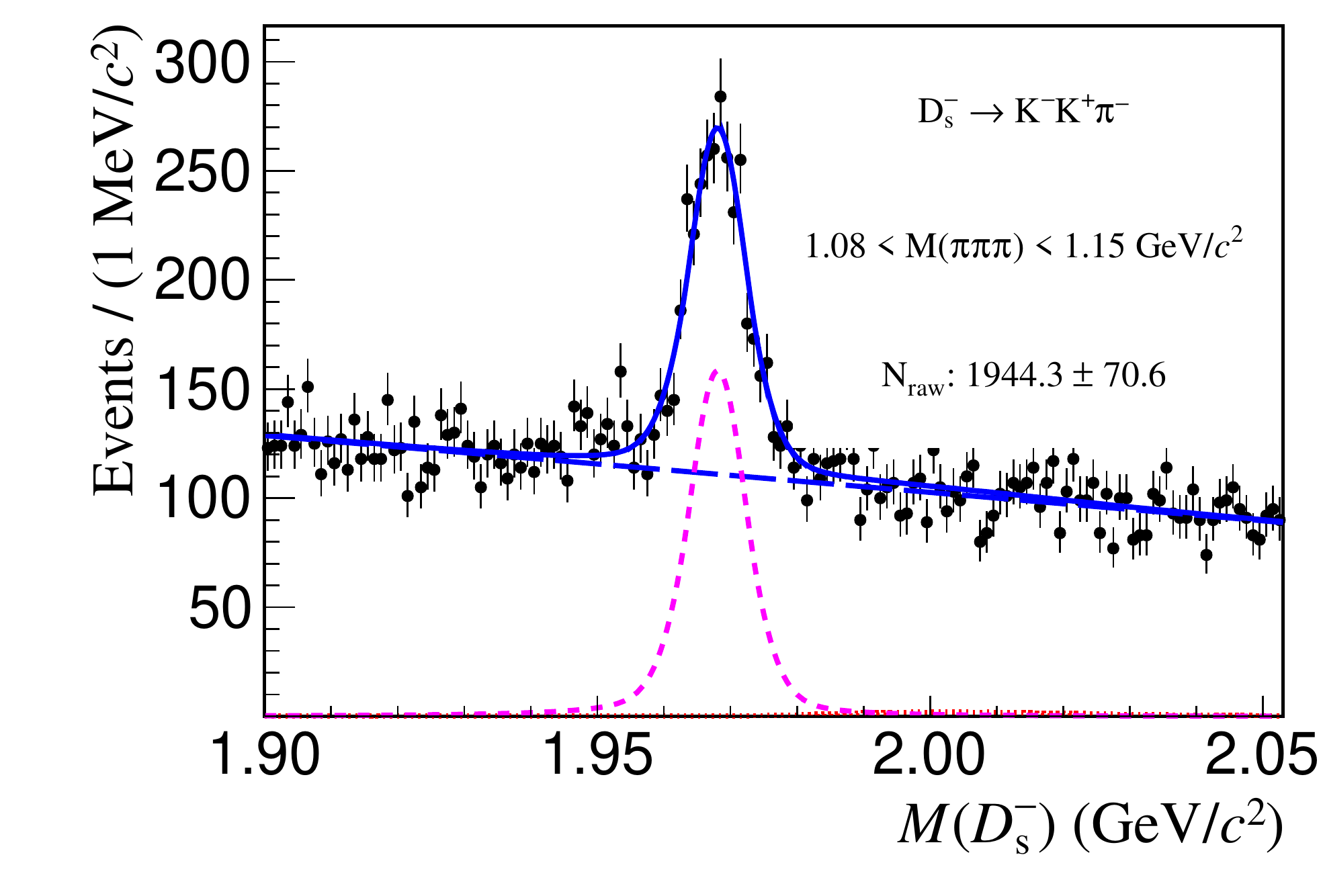}\\
\includegraphics[width=0.48\linewidth]{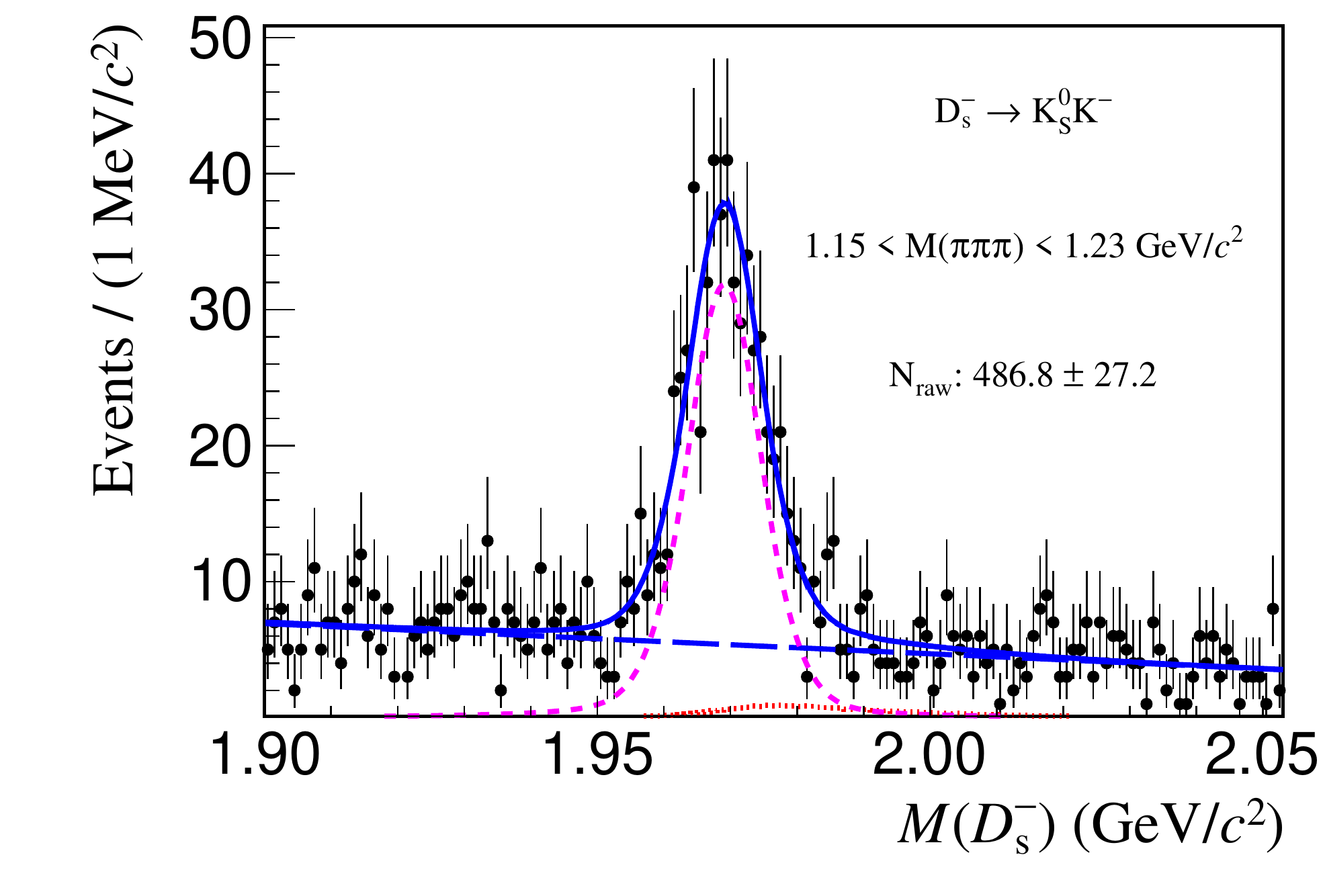}
\includegraphics[width=0.48\linewidth]{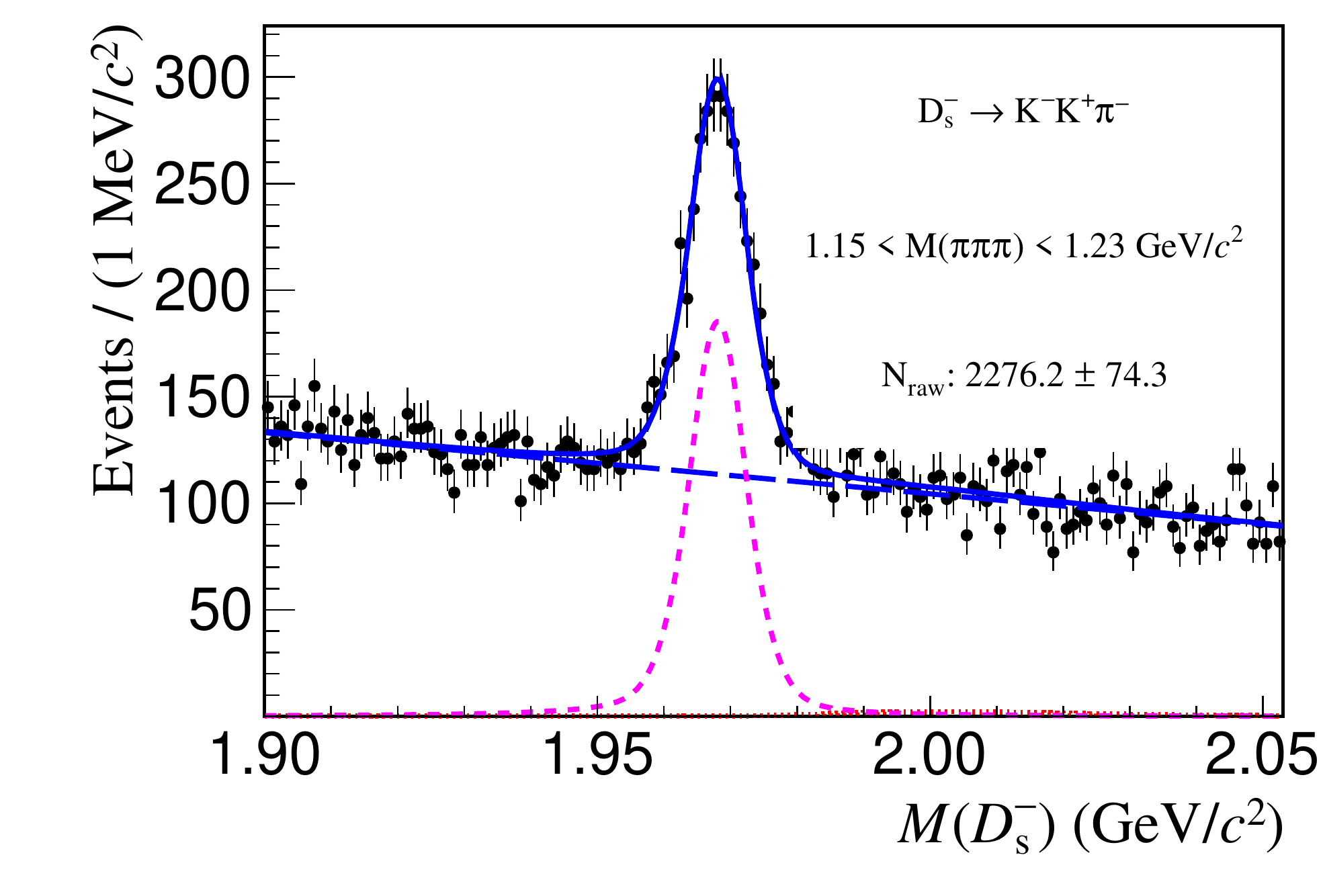}\\
\includegraphics[width=0.48\linewidth]{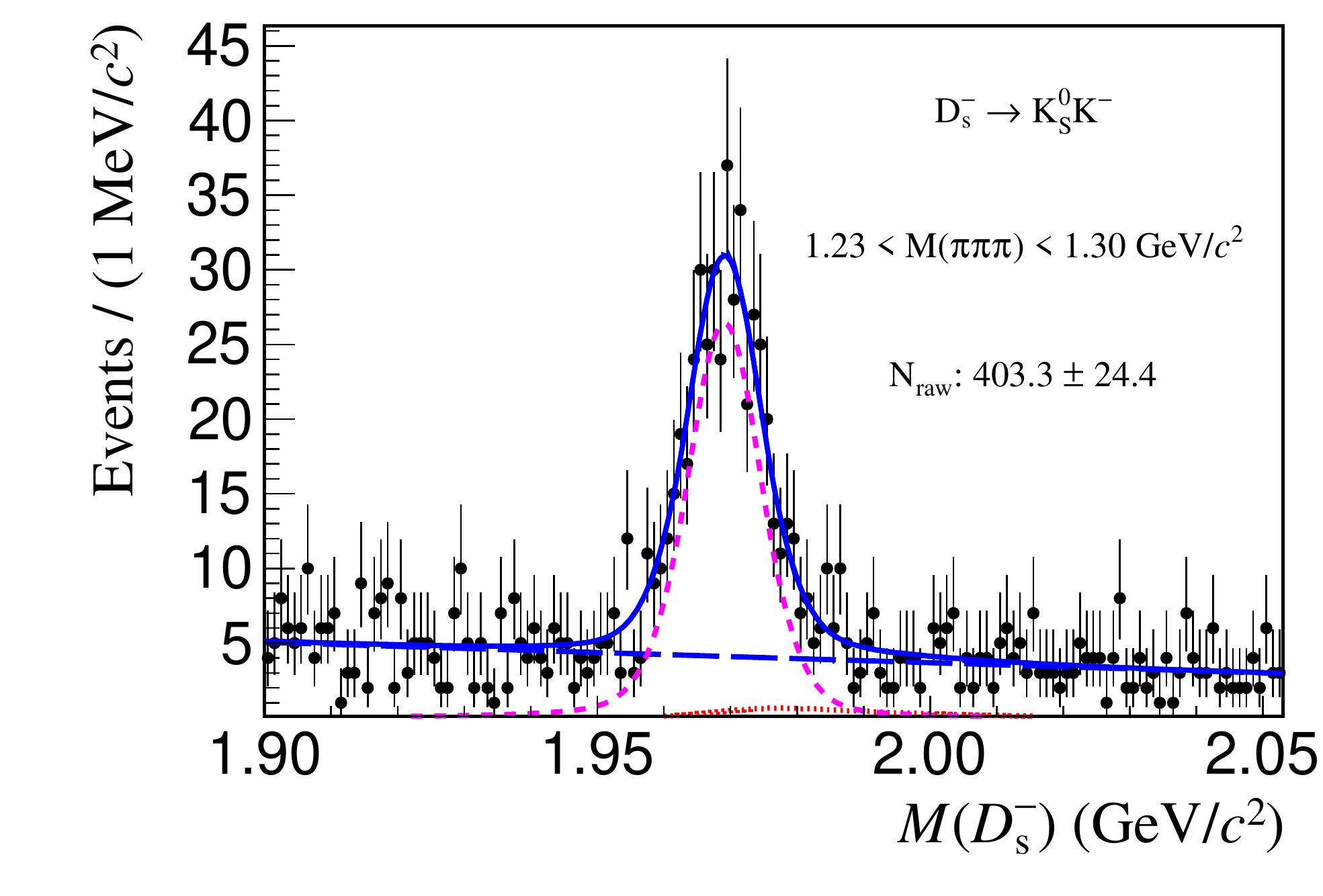}
\includegraphics[width=0.48\linewidth]{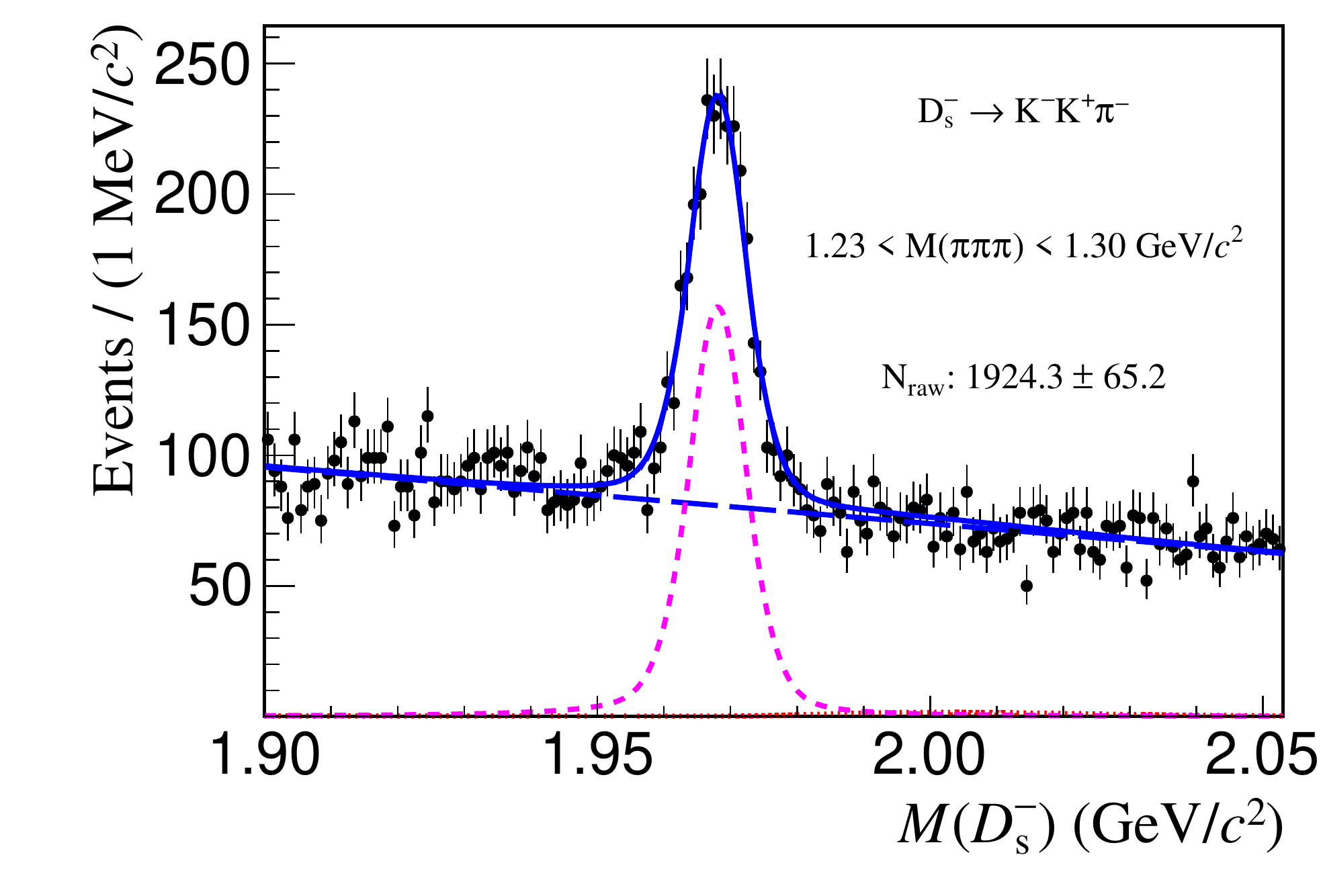}\\
\includegraphics[width=0.48\linewidth]{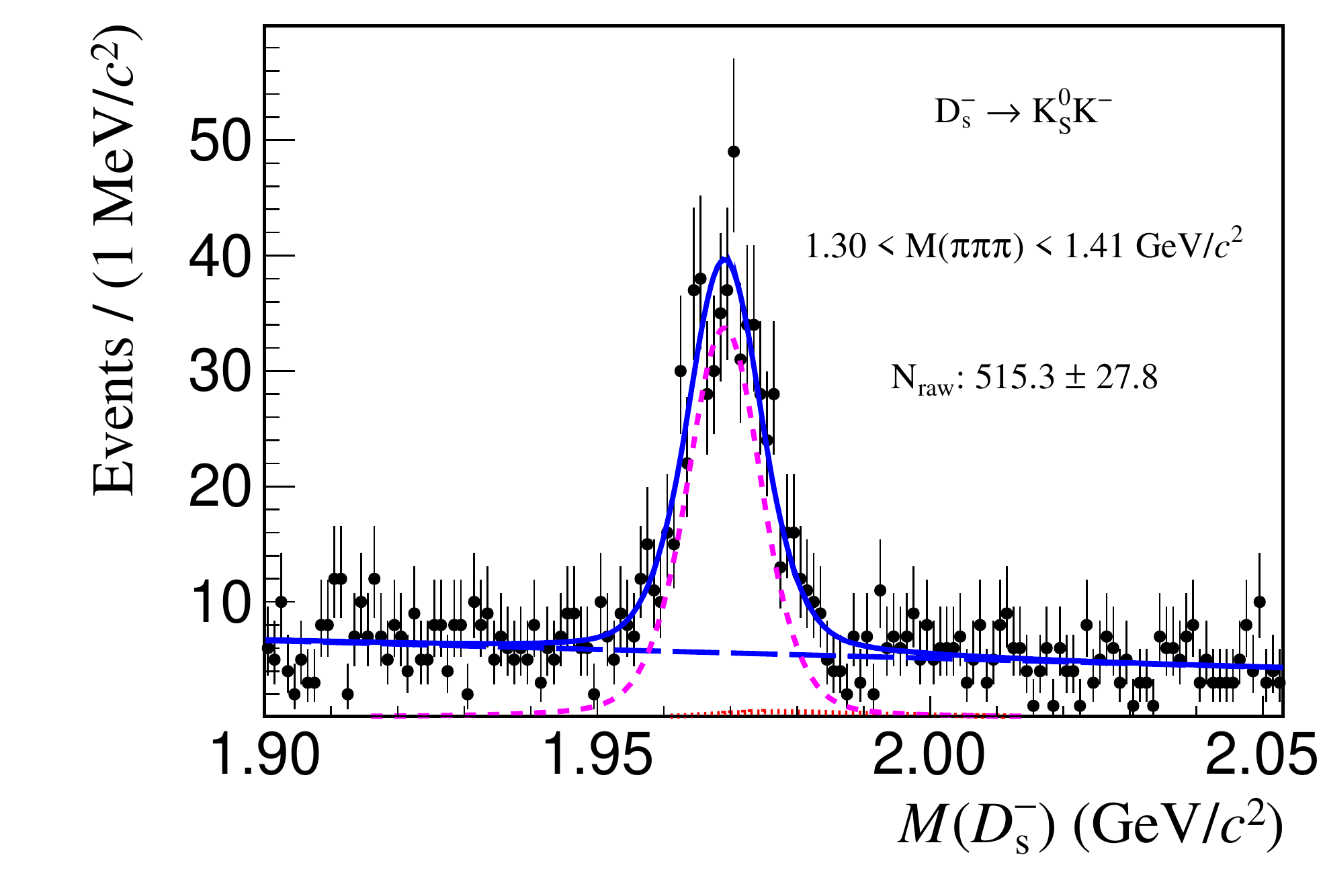}
\includegraphics[width=0.48\linewidth]{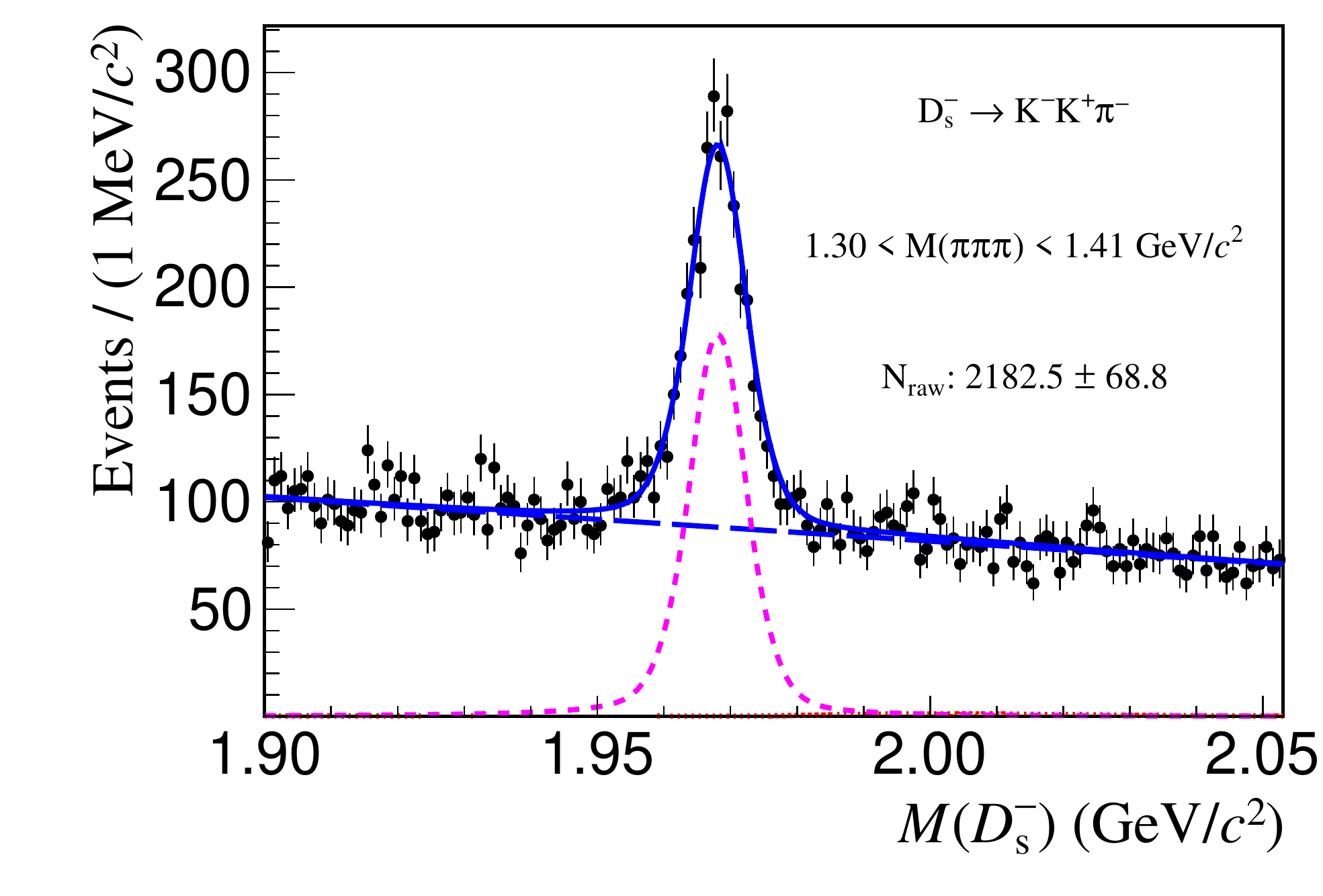}
%\begin{quote}
    \caption{The invariant mass $M(\Dsm)$ distributions %of the ST signals ($N^{\alpha}_{\rm ST}$) of 
    for the ST modes $\Dsm\to K_S^0K^-$ (left) and $\Dsm\to K^- K^+ \pi^-$ (right), part 2/3. Data (black points) are shown
    overlaid with the fit results including the total (solid blue),
signal PDF (dashed magenta), $\Dm$ background PDF (dotted red), and combinatorial background PDF (long-dashed blue) components.}
\label{fig:DTfitsvsm3pi_p2}
%\end{quote}
\end{figure}
\begin{figure}[!htb]
\centering
\includegraphics[width=0.48\linewidth]{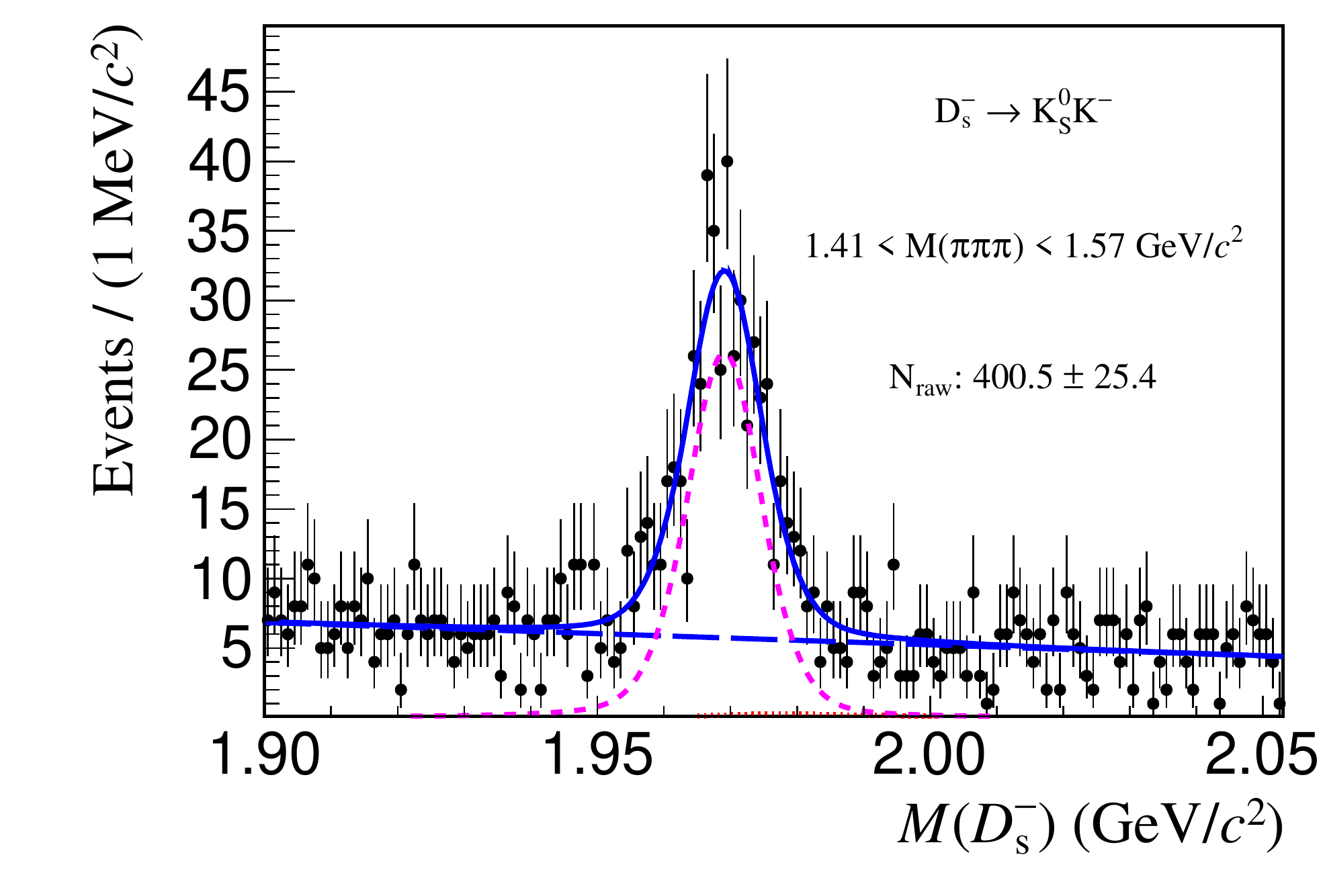}
\includegraphics[width=0.48\linewidth]{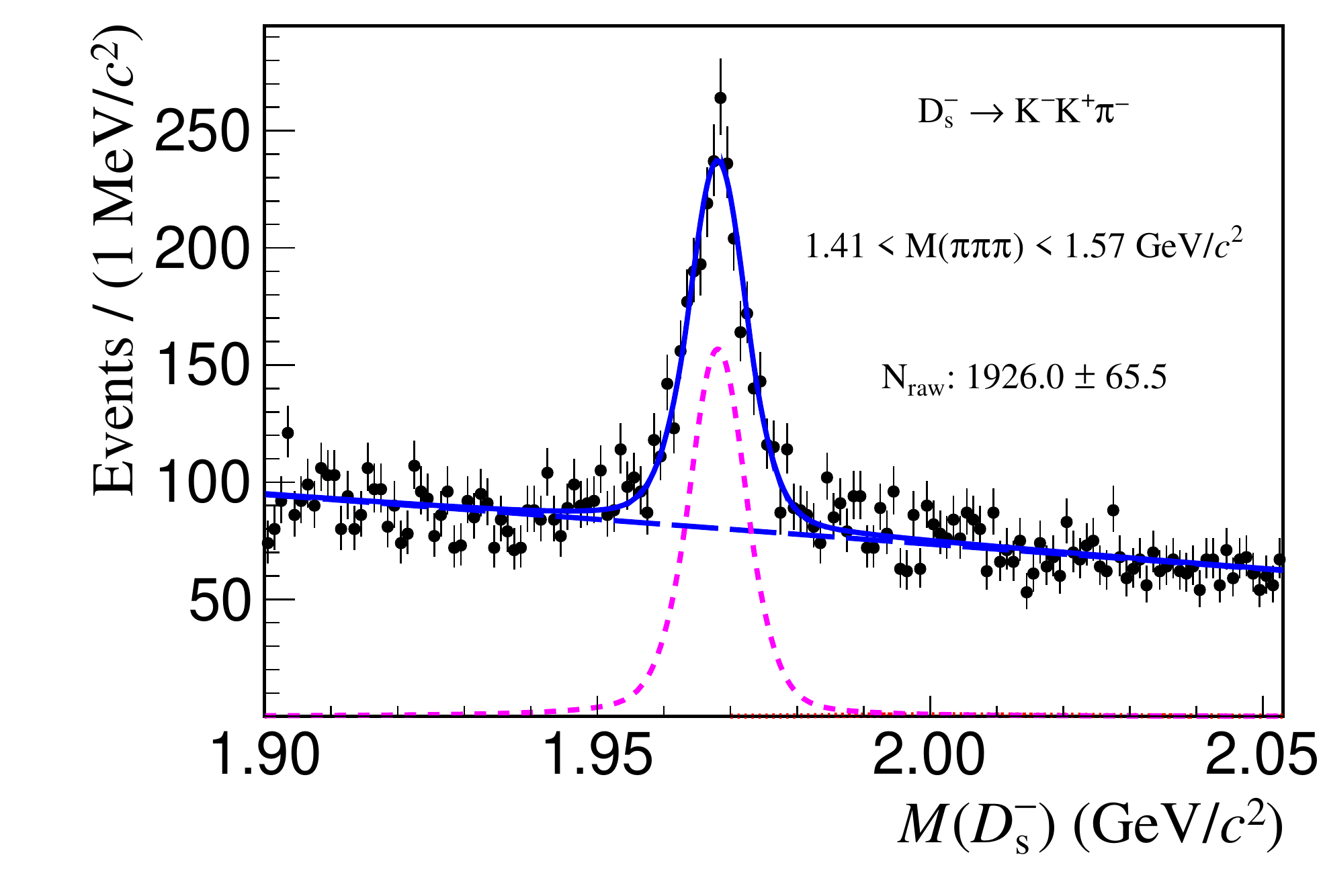}\\
\includegraphics[width=0.48\linewidth]{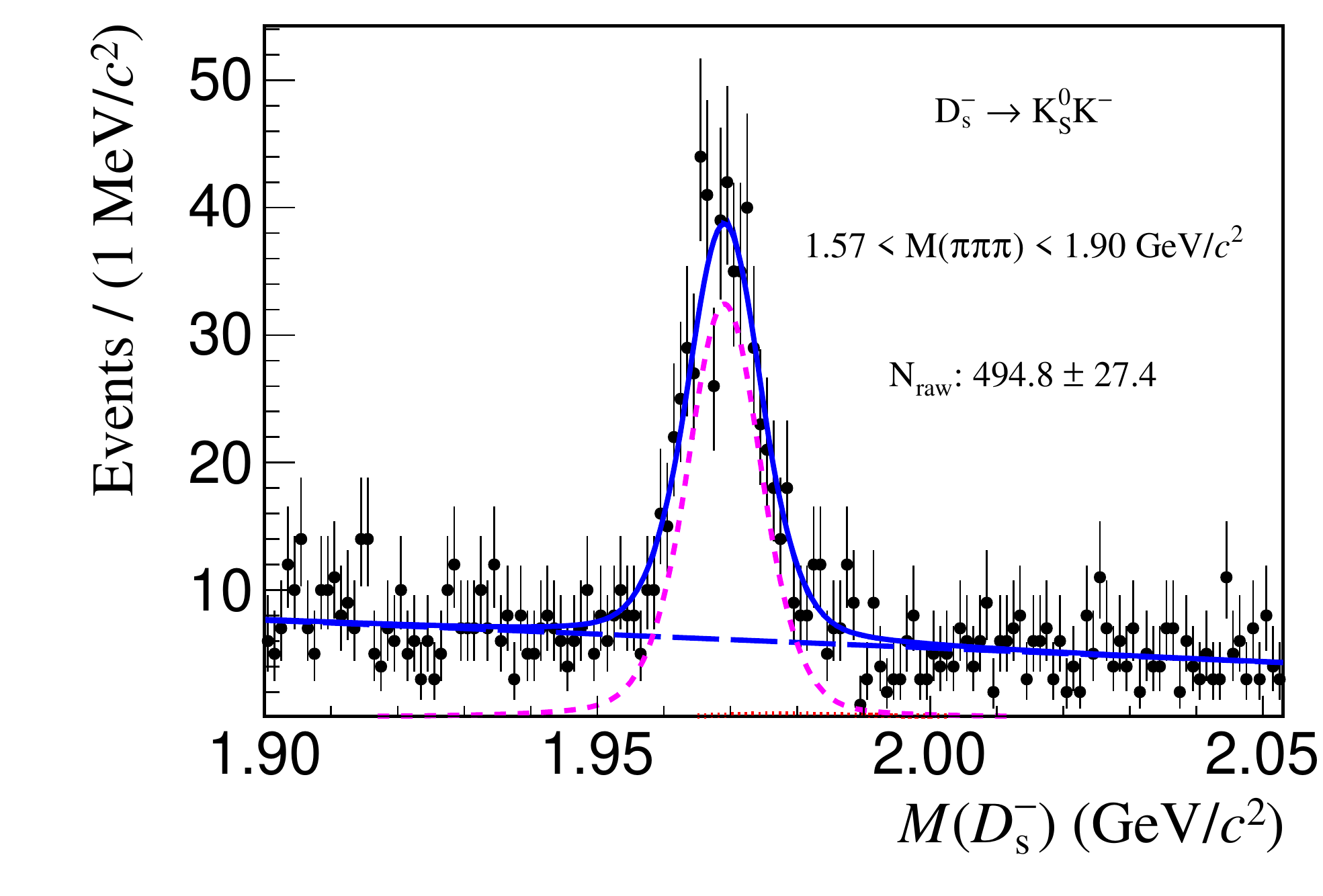}
\includegraphics[width=0.48\linewidth]{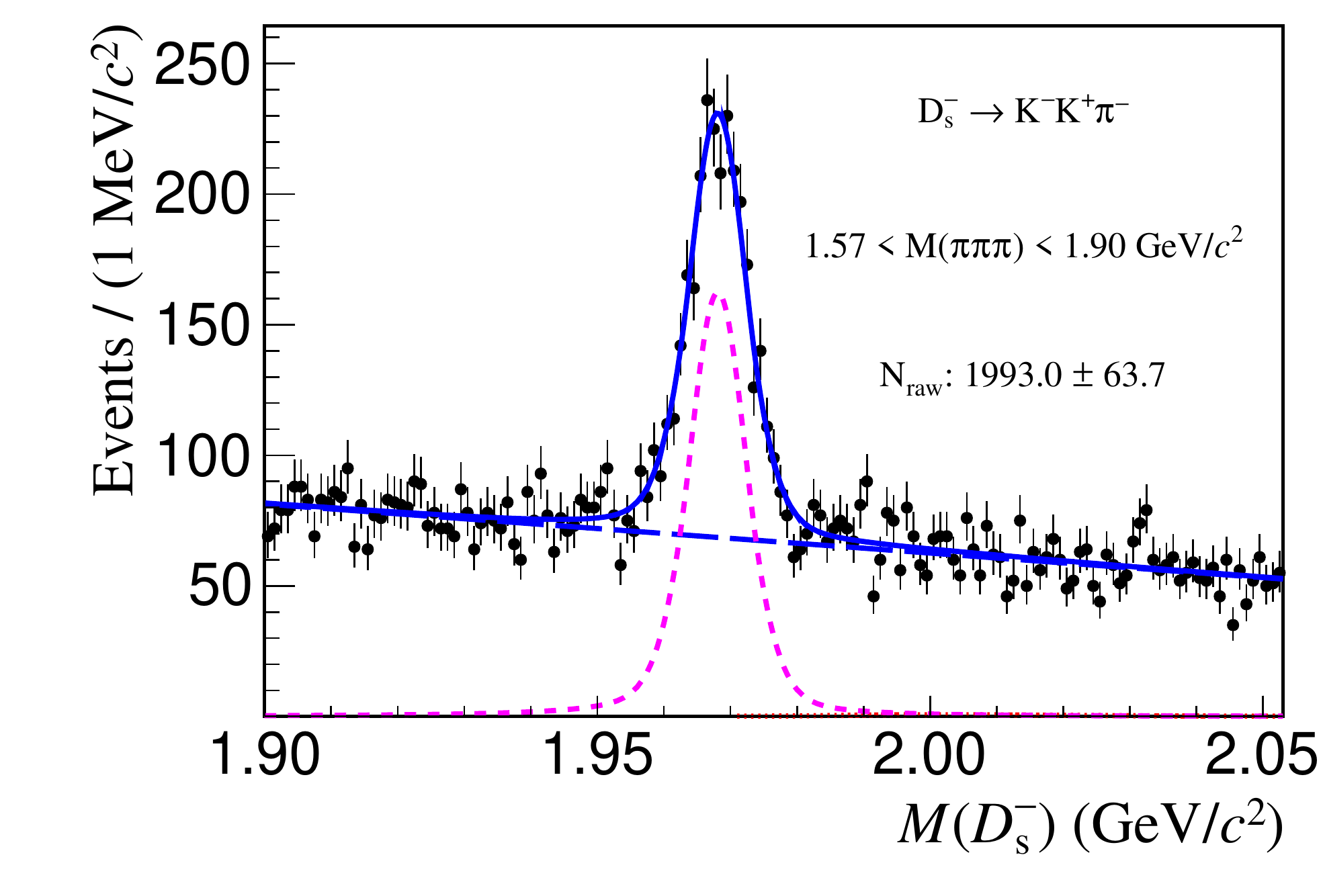}\\
\includegraphics[width=0.48\linewidth]{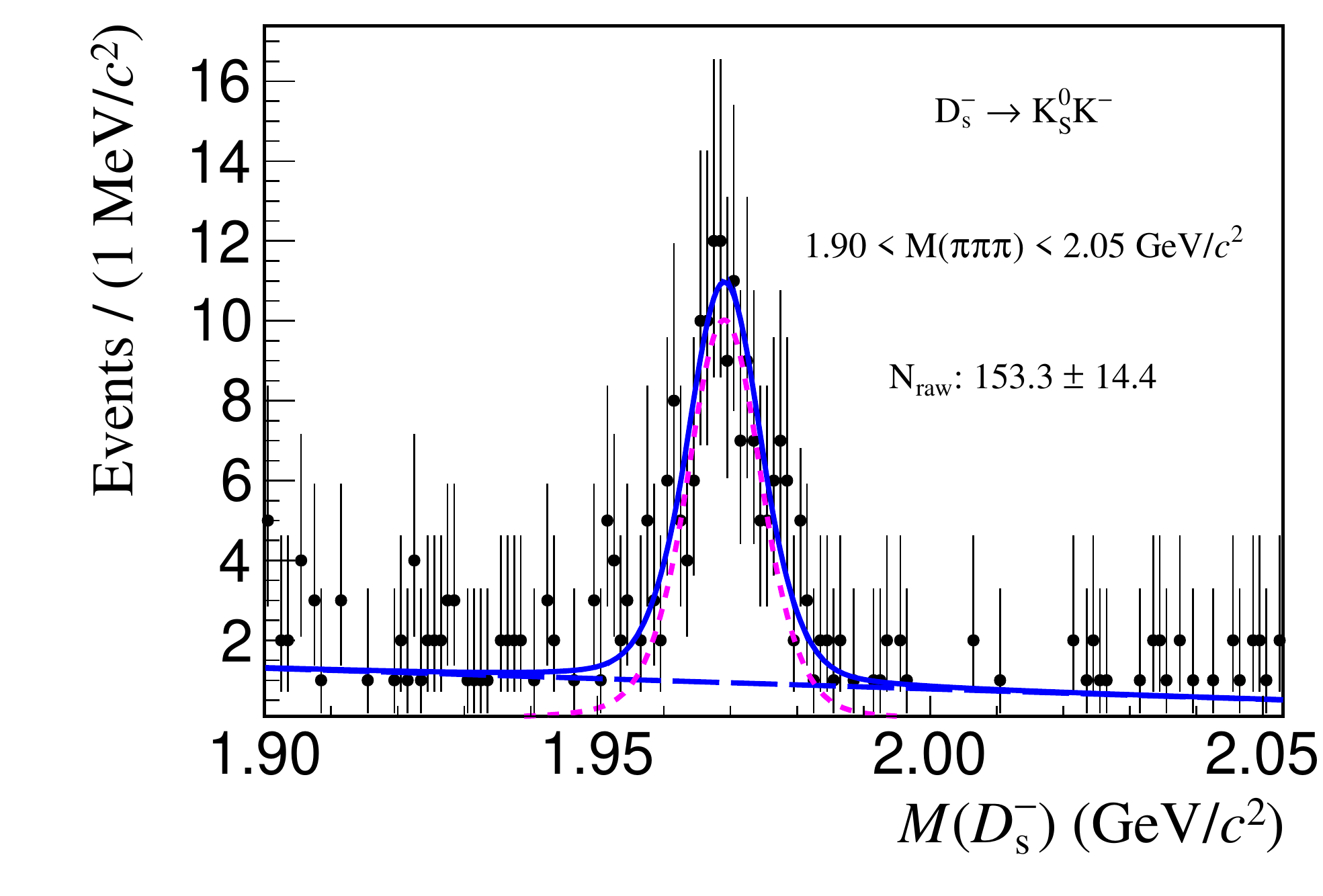}
\includegraphics[width=0.48\linewidth]{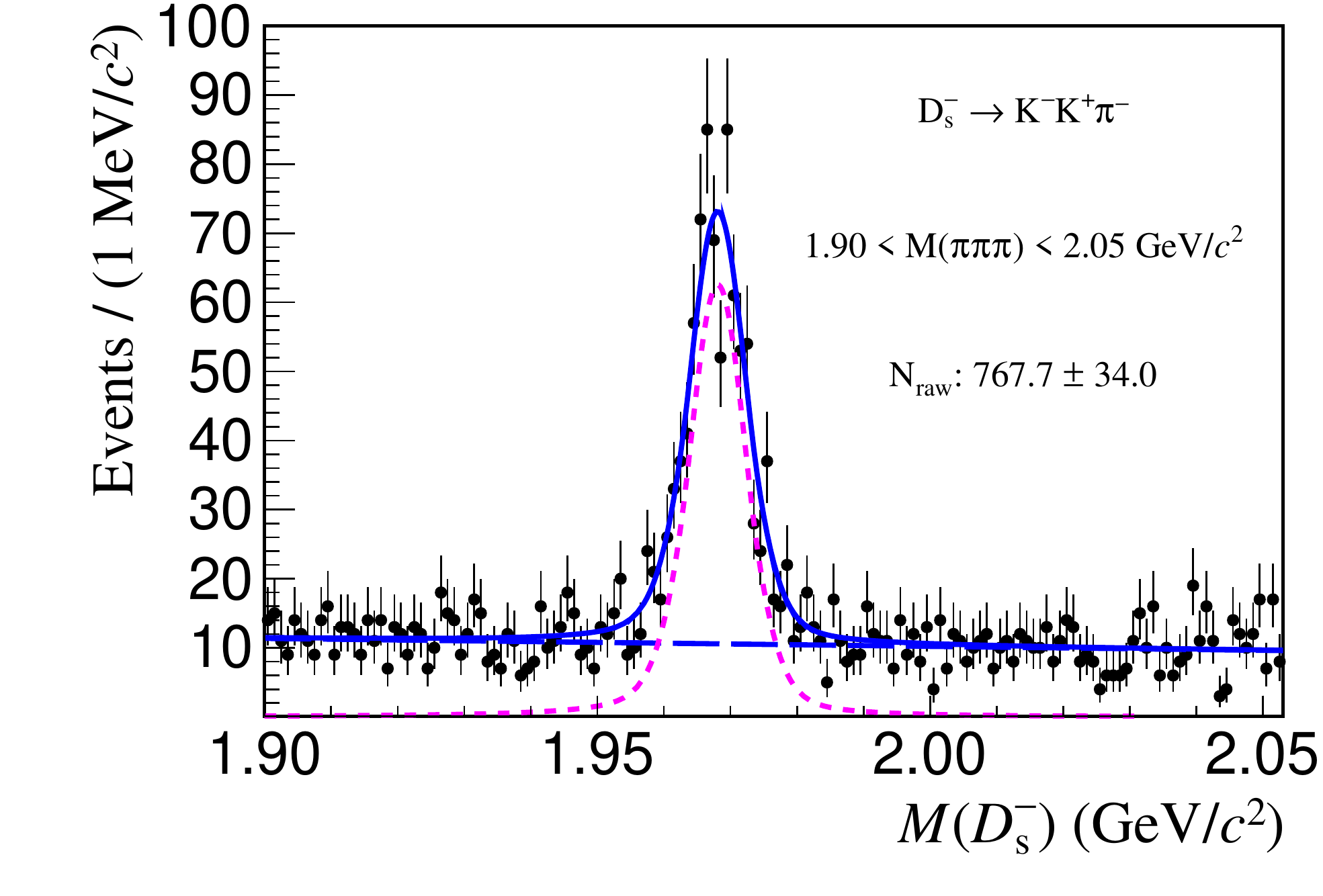}\\
%\begin{quote}
    \caption{The invariant mass $M(\Dsm)$ distributions %of the ST signals ($N^{\alpha}_{\rm ST}$) of 
    for the ST modes $\Dsm\to K_S^0K^-$ (left) and $\Dsm\to K^- K^+ \pi^-$ (right), part 3/3. Data (black points) are shown
    overlaid with the fit results including the total (solid blue),
signal PDF (dashed magenta), $\Dm$ background PDF (dotted red), and combinatorial background PDF (long-dashed blue) components.}
\label{fig:DTfitsvsm3pi_p3}
%\end{quote}
\end{figure}

\begin{table}[!htb]
\centering
    \caption{Summary of all different sources of systematic uncertainties for partial branching fractions of $D_s^+\to\pipipi X$. The uncertainties are summed up in quadrature to make the total systematic uncertainties.}
\label{tab:finalsys_bins}
%\footnotesize
\begin{tabular}{l|ccccccccccc}
\hline\hline
& \multicolumn{11}{c}{Relative uncertainty (\%) in $M(\pipipi)$ interval}\\
    Source & 1 & 2 &3 & 4 & 5 & 6 & 7 & 8 & 9 & 10 & 11\\ \hline
Sig. shape & 0.70  & 1.73  & 0.49  & 0.86  & 1.84  & 2.62  & 2.55  & 2.41  & 1.55  & 1.56  & 0.17 \\
Bkg. shape & 0.04  & 0.42  & 0.04  & 0.33  & 0.68  & 0.63  & 0.09  & 0.20  & 1.28  & 0.65  & 0.15 \\
%Tag bias & 0.03  & 0.03  & 0.03  & 0.03  & 0.03  & 0.03  & 0.03  & 0.03  & 0.03  & 0.03  & 0.03 \\
Track eff. & 1.50  & 1.50  & 1.50  & 1.50  & 1.50  & 1.50  & 1.50  & 1.50  & 1.50  & 1.50  & 1.50 \\
PID eff. & 1.50  & 1.50  & 1.50  & 1.50  & 1.50  & 1.50  & 1.50  & 1.50  & 1.50  & 1.50  & 1.50 \\
$K_S^0$ bkg. & 0.30  & 0.64  & 0.58  & 0.40  & 0.29  & 0.24  & 0.17  & 0.19  & 0.19  & 0.06  & 0.02 \\
Kaon misID & 0.40  & 0.49  & 0.42  & 0.36  & 0.30  & 0.23  & 0.17  & 0.14  & 0.19  & 0.28  & 0.02 \\
Lepton misID & 0.28  & 0.45  & 0.37  & 0.35  & 0.38  & 0.34  & 0.31  & 0.37  & 0.44  & 0.25  & 0.02 \\
Pion mult. & 0.91  & 1.12  & 1.11  & 0.73  & 0.39  & 0.19  & 0.01  & 0.15  & 0.33  & 0.06  & 0.00 \\
Pion mom. &  0.46  & 1.84  & 1.26  & 1.75  & 1.30  & 1.08  & 0.90  & 0.65  & 0.14  & 0.79  & 2.80 \\
MC size & 0.32  & 0.30  & 0.39  & 0.41  & 0.61  & 0.51  & 0.61  & 0.55  & 0.67  & 0.63  & 1.45 \\
Fit bias & 0.90  & 1.24  & 0.77  & 0.08  & 0.38  & 0.45  & 0.93  & 0.57  & 0.15  & 0.81  & 2.57 \\
\hline
Total  & 2.69  & 3.85  & 2.99  & 3.09  & 3.32  & 3.70  & 3.64  & 3.41  & 3.07  & 3.03  & 4.60 \\
\hline
\hline
\end{tabular}
\end{table}

%% file: authorlist_2022-08-25.tex
%% Saved at => 2022-08-25
\begin{small}
\begin{center}

M.~Ablikim$^{1}$, M.~N.~Achasov$^{12,b}$, P.~Adlarson$^{72}$, M.~Albrecht$^{4}$, R.~Aliberti$^{33}$, A.~Amoroso$^{71A,71C}$, M.~R.~An$^{37}$, Q.~An$^{68,55}$, Y.~Bai$^{54}$, O.~Bakina$^{34}$
% ...
\\
\vspace{0.2cm}
(BESIII Collaboration)\\
\vspace{0.2cm} {\it
$^{1}$ Institute of High Energy Physics, Beijing 100049, People's Republic of China\\
% ...
$^{4}$ Bochum  Ruhr-University, D-44780 Bochum, Germany\\
% ...
$^{12}$ G.I. Budker Institute of Nuclear Physics SB RAS (BINP), Novosibirsk 630090, Russia\\
% ...
$^{33}$ Johannes Gutenberg University of Mainz, Johann-Joachim-Becher-Weg 45, D-55099 Mainz, Germany\\
$^{34}$ Joint Institute for Nuclear Research, 141980 Dubna, Moscow region, Russia\\
% ...
$^{37}$ Liaoning Normal University, Dalian 116029, People's Republic of China\\
% ...
$^{54}$ Southeast University, Nanjing 211100, People's Republic of China\\
$^{55}$ State Key Laboratory of Particle Detection and Electronics, Beijing 100049, Hefei 230026, People's Republic of China\\
% ...
$^{68}$ University of Science and Technology of China, Hefei 230026, People's Republic of China\\
% ...
$^{71}$ University of Turin and INFN, (A)University of Turin, I-10125, Turin, Italy; (B)University of Eastern Piedmont, I-15121, Alessandria, Italy; (C)INFN, I-10125, Turin, Italy\\
$^{72}$ Uppsala University, Box 516, SE-75120 Uppsala, Sweden\\
% ...
$^{b}$ Also at the Novosibirsk State University, Novosibirsk, 630090, Russia\\
% ...
}\end{center}

\vspace{0.4cm}
\end{small}

%% file: main.bbl
\begin{thebibliography}{99}
\bibitem{hflav} 
	Y.~S.~Amhis \textit{et al.} (HFLAV Collaboration), \href{https://link.springer.com/article/10.1140/epjc/s10052-020-8156-7}{Eur. Phys. J. C \textbf{81}, 226 (2021)}; 
%
			Updated results available at \url{https://hflav-eos.web.cern.ch/hflav-eos/semi/winter23_prel/html/RDsDsstar/RDRDs.html}.

\bibitem{rdstlhcbprl} 
	R.~Aaij \textit{et al.} (LHCb Collaboration), \href{https://journals.aps.org/prl/abstract/10.1103/PhysRevLett.120.171802}{Phys. Rev. Lett. \textbf{120}, 171802 (2018)}; 
		R.~Aaij \textit{et al.} (LHCb Collaboration), \href{https://journals.aps.org/prd/abstract/10.1103/PhysRevD.97.072013}{Phys. Rev. D \textbf{97}, 072013 (2018)}.

%
%
%
%

\bibitem{LHCb:2022piu} R.~Aaij \textit{et al.} (LHCb Collaboration), \href{https://journals.aps.org/prl/abstract/10.1103/PhysRevLett.128.191803}{Phys. Rev. Lett. \textbf{128}, 191803 (2022)}.

\bibitem{pdg}
	R.~L. Workman {\it  et al.} (Particle Data Group), {Prog. Theor. Exp. Phys. \textbf{2022}, 083C01 (2022)}.
	%

\bibitem{besiii_etaprrhop} M. Ablikim {\it et al.} (BESIII Collaboration), %
	\href{https://doi.org/10.1016/j.physletb.2015.09.059}{Phys. Lett. B \textbf{750}, 466 (2015)}.

\bibitem{ds2eta3pi} M. Ablikim {\it et al.} (BESIII Collaboration), %
	\href{https://journals.aps.org/prd/abstract/10.1103/PhysRevD.104.L071101}{Phys. Rev. D \textbf{104}, L071101 (2021)}.

	\bibitem{BESIIIkkpipipi} M. Ablikim {\it et al.} (BESIII Collaboration), %
		\href{https://link.springer.com/article/10.1007/JHEP07(2022)051}{J. High Energy Phys. \textbf{07}, 051 (2022)}.

	\bibitem{BESIIIkkpi} M. Ablikim {\it et al.} (BESIII Collaboration), \href{https://journals.aps.org/prd/abstract/10.1103/PhysRevD.104.012016}{Phys. Rev. D \textbf{104}, 012016 (2021)}.

\bibitem{BESIIIkkpipi0} M. Ablikim {\it et al.} (BESIII Collaboration), \href{https://journals.aps.org/prd/abstract/10.1103/PhysRevD.104.032011}{Phys. Rev. D \textbf{104}, 032011 (2021)}.

\bibitem{BESIIItaunu} M. Ablikim {\it et al.} (BESIII Collaboration), \href{https://journals.aps.org/prd/abstract/10.1103/PhysRevD.104.052009}{Phys. Rev. D \textbf{104}, 052009 (2021)}; 
	M. Ablikim {\it et al.} (BESIII Collaboration), \href{https://journals.aps.org/prd/abstract/10.1103/PhysRevD.104.032001}{Phys. Rev. D \textbf{104}, 032001 (2021)}; 
		M. Ablikim {\it et al.} (BESIII Collaboration), \href{https://journals.aps.org/prl/abstract/10.1103/PhysRevLett.127.171801}{Phys. Rev. Lett. \textbf{127}, 171801 (2021)}.

\bibitem{ABLIKIM2010345} 
	M.~Ablikim \textit{et al.} (BESIII Collaboration), \href{https://doi.org/10.1016/j.nima.2009.12.050}{Nucl. Instrum. Meth. A \textbf{614}, 345 (2010)}.
	%

\bibitem{bepcii}  
C.~Yu \textit{et al.}, \href{https://accelconf.web.cern.ch/ipac2016/doi/JACoW-IPAC2016-TUYA01.html}{ Proceedings of IPAC2016, Busan, Korea, 642 2016}.
	%

\bibitem{dataset} 
	M.~Ablikim \textit{et al.} (BESIII Collaboration), \href{https://iopscience.iop.org/article/10.1088/1674-1137/44/4/040001}{Chin. Phys. C \textbf{44}, 040001 (2020)}.
	%

\bibitem{besiii_mag} 
	K.~X.~Huang \textit{et al.}, \href{https://doi.org/10.1007/s41365-022-01133-8}{Nucl.\ Sci.\ Tech. \textbf{33}, 142 (2022)}.

\bibitem{detector2015} 
	X.~Li \textit{et al.}, \href{https://doi.org/10.1007/s41605-017-0014-2}{Radiat Detect Technol Methods \textbf{1}, 13 (2017)};
		Y.X. Guo \textit{et al.}, \href{https://doi.org/10.1007/s41605-017-0012-4}{Radiat. Detect. Technol. Methods \textbf{1}, 15 (2017)};
  P.~Cao \textit{et al.}, \href{https://doi.org/https://doi.org/10.1016/j.nima.2019.163053}{Nucl. Instrum. Meth. A \textbf{953}, 163053 (2020)}.
	%

\bibitem{sim} 
    S. Agostinelli {\it et al.} (GEANT4 collaboration), \href{https://doi.org/10.1016/S0168-9002(03)01368-8}{Nucl. Instrum. Meth. A \textbf{506}, 250 (2003)}.
	%

\bibitem{KKMC} 
	S.~Jadach, B.~F.~L.~Ward and Z.~Was, \href{https://journals.aps.org/prd/abstract/10.1103/PhysRevD.63.113009}{Phys. Rev. D \textbf{63}, 113009 (2001)}.
	%

\bibitem{EvtGen} 
	D.~J.~Lange, \href{https://doi.org/10.1016/S0168-9002(01)00089-4}{Nucl. Instrum. Meth. A \textbf{462}, 152 (2001)};
		R.~G.~Ping, \href{https://doi.org/10.1088/1674-1137/32/8/001}{Chin. Phys. C \textbf{32}, 599 (2008)}.
	%
   %

\bibitem{LundCharm} 
	J.~C.~Chen, G.~S.~Huang, X.~R.~Qi, D.~H.~Zhang and Y.~S.~Zhu, \href{https://journals.aps.org/prd/abstract/10.1103/PhysRevD.62.034003}{Phys. Rev. D \textbf{62}, 034003 (2000)};
	R.~L.~Yang, R.~G.~Ping and H.~Chen, \href{https://iopscience.iop.org/article/10.1088/0256-307X/31/6/061301}{Chin. Phys. Lett. \textbf{31}, 061301 (2014)}.
	%
	%

\bibitem{FSR} E. Richter-Was, 
	\href{https://doi.org/10.1016/0370-2693(93)90062-M}{Phys. Lett. B {\bf 303}, 163 (1993)}.

\bibitem{dtref} R. M. Baltrusaitis {\it et al.} (Mark III Collaboration), \href{https://journals.aps.org/prl/abstract/10.1103/PhysRevLett.56.2140}{Phys. Rev. Lett. {\bf 56}, 2140 (1986)}; J. Adler {\it et al}. (Mark III Collaboration), \href{https://journals.aps.org/prl/abstract/10.1103/PhysRevLett.60.89}{Phys. Rev. Lett. {\bf 60}, 89 (1988)}.    

\bibitem{ds2pn} M. Ablikim {\it et al.} (BESIII Collaboration), 
	\href{https://journals.aps.org/prd/abstract/10.1103/PhysRevD.99.031101}{Phys. Rev. D {\bf 99}, 031101 (2019)}.

\bibitem{dscb} T.~Skwarnicki, Ph.D.~thesis, Institute of Nuclear Physics, Krakow Poland, (1986).

\bibitem{ds23pi} M. Ablikim {\it et al.} (BESIII Collaboration), \href{https://journals.aps.org/prd/abstract/10.1103/PhysRevD.106.112006}{Phys. Rev. D \textbf{106}, 112006 (2022)}.  

\bibitem{barlow} R. Barlow, \href{https://arxiv.org/abs/hep-ex/0207026}{arXiv:hep-ex/0207026}.

\bibitem{bes3d2klnu} M. Ablikim {\it et al.} (BESIII Collaboration), \href{https://journals.aps.org/prl/abstract/10.1103/PhysRevLett.122.011804}{Phys. Rev. Lett. \textbf{122}, 011804 (2019)}.  

\bibitem{BESIIIDs2eX} M.~Ablikim \textit{et al.} (BESIII Collaboration), 
	\href{https://journals.aps.org/prd/abstract/10.1103/PhysRevD.104.012003}{Phys. Rev. D \textbf{104}, 012003 (2021)}.

\bibitem{supp} See Supplemental Material at \url{http://link.aps.org/supplemental/10.1103/PhysRevD.108.032001}, for the full set of fits used in determining the DT yields.


\bibitem{gbrw} 
A.~Rogozhnikov, \href{https://iopscience.iop.org/article/10.1088/1742-6596/762/1/012036}{J. Phys. Conf. Ser. \textbf{762}, 012036 (2016)}.

	\bibitem{lhcb-white-paper}
R. Aaij {\it et al.} (LHCb Collaboration), \href{https://arxiv.org/abs/1808.08865}{[arXiv:1808.08865 [hep-ex]]}.

\end{thebibliography}
